\documentclass[11pt]{article}
\usepackage{geometry}
\usepackage{amsfonts,amsmath,amssymb}
\usepackage{mathrsfs,frcursive}
\usepackage{color}
\usepackage{theorem,cite}
\usepackage{hyperref}
\usepackage{graphicx, ulem}

\newtheorem{thm}{Theorem}[section]
\newtheorem{prop}[thm]{Proposition}

\newtheorem{conj}[thm]{Conjecture}

\newtheorem{rmk}[thm]{Remark}

\newcommand{\bea}{\begin{eqnarray}}
\newcommand{\eea}{\end{eqnarray}}
\newcommand{\beali}{\begin{align}}
\newcommand{\eeali}{\end{align}}
\newcommand{\beano}{\begin{eqnarray*}}
\newcommand{\eeano}{\end{eqnarray*}}
\newcommand{\beq}{\begin{equation}}
\newcommand{\eeq}{\end{equation}}

\newcommand{\mb}[1]{\hspace{2.1ex}\mbox{#1}\hspace{2.1ex}}

\def\fh{{\mathfrak h}}

\def\fn{{\mathfrak n}}

\def\fp{{\mathfrak p}}

\def\fr{{\mathfrak r}}
\def\fs{{\mathfrak s}}
\def\ft{{\mathfrak t}}

\def\cA{{\cal A}}    
    
    \def\cI{{\cal I}}
    \def\cL{{\cal L}}
    
\def\cP{{\cal P}}  \def\cQ{{\cal Q}}  
\def\cS{{\cal S}}  \def\cT{{\cal T}}

\newcommand{\CC}{{\mathbb C}}

\newcommand{\II}{{\mathbb I}}

\newcommand{\mr}{{\textsf{r}}}

\newcommand{\wt}[1]{\widetilde{#1}}

\newcommand{\sfrac}[2]{{\textstyle{\frac{#1}{#2}}}}

\newcommand{\proof}{\textbf{Proof:}~}
\newcommand{\qed}{{\hfill \rule{5pt}{5pt}}\\}

\def\eps{\varepsilon}

\def\sgn{\mathop{\rm sgn}\nolimits}

\def\tr{\mathop{\rm tr}\nolimits}

\newcommand\atopn[2]{\genfrac{}{}{0pt}{}{#1}{#2}}

\def\pp{|\!|p|\!|}

\newcommand{\doi}[1]{\href{https://doi.org/#1}{doi:#1}}

\numberwithin{equation}{section}

\begin{document}
\pagestyle{empty}

\null
\vspace{20pt}

\parskip=6pt

\begin{center}
\begin{LARGE}
\textbf{Quadratic Poisson brackets for\\[1ex]
the Camassa--Holm peakons}
\end{LARGE}
\end{center}

\begin{center}
J. Avan\textsuperscript{(a)}, 
L. Frappat\textsuperscript{(b)}, 
E. Ragoucy\textsuperscript{(b)} 
\footnote{avan@cyu.fr, luc.frappat@lapth.cnrs.fr (corresponding author), eric.ragoucy@lapth.cnrs.fr}
\end{center}

\begin{center}
{\bf (a)} Laboratoire de Physique Th\'eorique et Mod\'elisation, \\ 
CY Cergy Paris Universit\'e, CNRS, F-95302 Cergy-Pontoise, France
\\
{\bf (b)} Laboratoire d'Annecy-le-Vieux de Physique Th{\'e}orique LAPTh, \\ 
CNRS, Université Savoie Mont Blanc, F-74940 Annecy
\\
\end{center}


\vspace{4mm}

\subsection*{Abstract}
\textsl{
We establish quadratic Poisson brackets for the generalized Camassa--Holm peakon structure introduced in \cite{AFR23}. The calculation is based on the halving of the spectral parameter dependent $r$-matrix used to define the linear Poisson structure of this model. This quadratic structure, together with the linear one, establish the bi-Hamiltonian structure of the generalized Camassa--Holm peakon model.
\\
When the deformation parameter tends to $\pm2$, the spectral parameter dependence drops out, and we recover the linear and quadratic Poisson structure of the Camassa--Holm peakon model.
\\
When the spectral parameter tends to the fixed points of the involution defining the halving,  we recover the Ragnisco--Bruschi deformation of the Camassa--Holm peakon model, thereby establishing a new quadratic Poisson structure thereof.
\\[1ex]
{MSC classification: 35Q51 ; 37K10 ; 37K20}
}


\section{Introduction\label{sec:intro}}

The Camassa--Holm \cite{CH} peakons are a remarkable example of $N$-body integrable system deduced from a 1+1 dimensional integrable fluid equation in a way which preserves classical integrability. Peakon solutions exhibit a singular behaviour
of the form 
\begin{equation}
u(x, t) = \sum_{i=1}^N p_i(t)\, e^{\vert x-q_i(t) \vert\,. }
\end{equation}
Integrability of the Camassa--Holm peakons (hereafter denoted as ``strict peakons'') was established in \cite{CH,CHH} stemming from the canonical Poisson structure for the variables $(p_i, q_j)$. A one-parameter extension was proposed by Ragnisco and Bruschi \cite{RB}. However integrability under the Lax formulation proposed by these authors had a rather awkward formulation in terms of an implicit dynamical classical $r$ matrix, strongly suggesting that a more self contained formulation should exist. This was achieved in our paper \cite{AFR23} where a Lax matrix depending on a spectral parameter and a deformation parameter was proposed, together with  a linear $r$-matrix formulation i.e.:
\begin{equation}
\{L_1, L_2\} = [r_{12}, L_1] - [r_{21}, L_2]\,.
\end{equation}
It provided a more general peakon dynamics which in turns yielded a new peakon-type solution of the Camassa–Holm equation.

A remarkable feature of integrable peakons is their multi-Hamiltonian formulation (see e.g. \cite{Magri78,OR}). It is indeed possible to reformulate the dynamics of such integrable peakons using a second Poisson structure for $p_i, q_j$ with such features as dynamical dependence of the Poisson brackets (i.e. in the variables themselves). This was achieved in our paper \cite{AFR22}.

\paragraph{Main result.}
We address and solve in this paper the question of a multi-Hamiltonian formulation for the new generalized integrable peakons 
introduced in \cite{AFR23}.  Precisely we will derive a Poisson structure (albeit for a slightly restricted set of dynamical variables $(p_i, q_{jk} \equiv q_j-q_k)$ excluding the center of mass), dynamical in the variables, and inducing a natural quadratization of the linear structure found in \cite{AFR23}. 
The construction relies on a $r$-matrix $r_{12}(z_1,z_2)$, an idempotent morphism which is the composition of the transposition with a birational mapping $\sigma$ acting on the spectral parameter $z$, and a Lax matrix $\cL(z)$.
The two Poisson structures are given by 
\begin{subequations}\label{eq:PB-structures}
\begin{eqnarray}
&&\begin{split}
\big\{\cL_1(z_1)\,,\,\cL_2(z_2)\big\}_{(1)} =\ & \big[ r_{12}(z_1,z_2)+r_{12}(z_1,z^\sigma_2)^{t_2}\,,\,\cL_1(z_1)\big]\\
&-\big[ r_{21}(z_2,z_1)+r_{21}(z_2,z^\sigma_1)^{t_1}\,,\,\cL_2(z_2)\big]\,,
\end{split}\label{PB:lin}
\\
&&\begin{split}
\big\{\cL_1(z_1)\,,\,\cL_2(z_2)\big\}_{(2)} =\ &\big[ r_{12}(z_1,z_2)\,,\,\cL_1(z_1)\,\cL_2(z_2)\big]\\
&+\cL_1(z_1)\,r_{12}(z^\sigma_1,z_2)^{t_1}\,\cL_2(z_2) -\cL_2(z_2)\,r_{12}(z^\sigma_1,z_2)^{t_1}\,\cL_1(z_1)\,.
\end{split}\label{PB:quad}
\end{eqnarray}
\end{subequations}
The linear Poisson structure  \eqref{PB:lin} was described in \cite{AFR23}, and provides a generalization of the linear Ragnisco--Bruschi structure given in \cite{RB}. The quadratic Poisson structure  \eqref{PB:quad} is new, both for the Ragnisco--Bruschi model and for its generalization presented in \cite{AFR23}.
These two Poisson structures \eqref{eq:PB-structures} are compatible, implying a bi-Hamiltonian structure for the generalized peakon model.

The plan runs as follows:

We recall in Section \ref{sec:PB} the original construction of the generalized Camassa--Holm peakons in \cite{AFR23}, and the reduction to the strict peakon Lax matrix formulation for the canonical Poisson structure. In Section \ref{sec:split_alg} we identify our structure as the result of a ``folding'' of another Lax representation  denoted as ``halved algebra'' for reasons which will clearly appear.  A quadratic Poisson structure is also proposed  for the halved algebra. In Section  \ref{sec:quad-peakon} we identify the expected form of a quadratic $r$-matrix structure for a Lax matrix obtained by folding of an initial object on the lines of Section \ref{sec:split_alg}. We then compute the Poisson structure for the variables $(p_i, q_{jk} \equiv q_j-q_k)$ allowing to realize this quadratic structure from the giving of a Lax matrix and the $r$-matrix. 
As a by-product, we derive a quadratic structure for the Ragnisco–Bruschi extension of the Camassa–Holm peakons, and we again establish a link with the quadratic form of the strict peakon case. 
Section \ref{sec:hamil} is devoted to the dynamics associated to this quadratic structure. We compute the first Hamiltonians and determine the corresponding time evolution. Then, we present a conjecture on the general form of the transfer matrix and determine the cyclic variable appearing in the problem.
We conclude in Section \ref{sec:conclu} on open problems. The appendix \ref{app:hamil} describes the Hamiltonian dynamics for the halved algebra, while two other appendices present the technical derivation of Poisson brackets for the peakon variables (appendix~\ref{appB}) and provide explicit formulas for selected transfer matrices (appendix~\ref{appC}).

\section{Poisson structure of the Camassa--Holm peakon model and its generalization\label{sec:PB}}
We recall the spectral parameter formulation of the generalized Camassa--Holm peakons \cite{AFR23}. It is based on the Poisson algebra of three building blocks, namely three matrices $A$, $T$, and $S$. Their Poisson brackets are determined by the canonical Poisson structure of dynamical variables $p_i$ and $q_i$, $i=1,...,N$. The generalization also depends on a deformation parameter $\gamma$.
We then detail how a specific limit of this deformation parameter yields the original peakon model, while the study of the fixed points of a certain involution naturally leads to the Ragnisco–Bruschi model.
\subsection{The peakon algebra}
We recall here the Poisson structure of the Lax representation associated to the deformation of the Camassa--Holm peakon model.
For such a purpose we introduce the matrices
\begin{align}
A &= \sum_{i,j=1}^N \sqrt{p_ip_j} \,\sinh \frac{\nu}{2}(q_i-q_j) \, E_{ij} \label{eq:A}\,,\\
T &= \sum_{i,j=1}^N \sqrt{p_ip_j} \,\cosh \frac{\nu}{2}(q_i-q_j) \, E_{ij} \label{eq:T}\,,\\
S &= \sum_{i,j=1}^N \sqrt{p_ip_j} \,\fs_{ij}\,\sinh \frac{\nu}{2}(q_i-q_j) \, E_{ij}\,, \label{eq:S}
\end{align}
where $\fs_{ij}=\sgn(q_i-q_j) $ with the convention $\fs_{ii}=0$.
Then, the canonical Poisson structure for the peakon variables 
\begin{equation}\label{eq:PBcano}
\big\{p_i\,,\,p_j\big\}_{(1)}  =0\,,\quad \big\{q_i\,,\,p_j\big\}_{(1)} =\delta_{ij}\,,\quad \big\{q_i\,,\,q_j\big\}_{(1)} =0\,,
\end{equation} 
leads to the following Poisson brackets
\begin{subequations}\label{eq:PB}
\begin{equation}
\begin{split}
\big\{ A_1 , A_2 \big\}_{(1)} &= \frac{\nu}{4}\;\big[ \Pi-\Pi^{t},A_1 \big] \quad\ ; \qquad 
\big\{ T_1 , A_2 \big\}_{(1)} = \frac{\nu}{4}\;\big[ \Pi-\Pi^{t},T_1 \big]\,,\\
\big\{ A_1 , T_2 \big\}_{(1)} &= \frac{\nu}{4}\;\big[ \Pi+\Pi^{t},T_1 \big]  \quad\ ; \qquad
\big\{ T_1 , T_2 \big\}_{(1)} = \frac{\nu}{4}\;\big[ \Pi+\Pi^{t},A_1 \big]\,,\\
\big\{ A_1 , S_2 \big\}_{(1)} &= -\frac{\nu}{4}\;\big[ {\Gamma}_{12},A_1 \big]  \qquad ; \qquad
\big\{ T_1 , S_2 \big\}_{(1)} = -\frac{\nu}{4}\;\big[ {\Gamma}_{12},T_1 \big] \,,
\label{eq:PB-AT1}
\end{split}
\end{equation}
\begin{equation}
\begin{split}
\big\{ S_1 , A_2 \big\}_{(1)} &= \frac{\nu}{4}\;\big[ {\Gamma}_{21},A_2 \big]\,, \\
\big\{ S_1 , T_2 \big\}_{(1)} &= \frac{\nu}{4}\;\big[ {\Gamma}_{21},T_2 \big] \,,\\ 
\big\{ S_1 , S_2 \big\}_{(1)} &= -\frac{\nu}{4}\;\left( \big[ \Pi+\Pi^{t},A_1 \big] + \big[ {\Gamma}_{12},S_1 \big] - \big[ {\Gamma}_{21},S_2 \big] \right) \,,
\label{eq:PB-S1}
\end{split}
\end{equation}
\end{subequations}
with 
\begin{equation}\label{eq:defT}
\begin{split}
&\Pi=\sum_{i,j=1}^N E_{ij}\otimes E_{ji} \mb{;} \Pi^t=\sum_{i,j=1}^N E_{ij}\otimes E_{ij}\,,
\\
& {\Gamma}_{12}=\sum_{i,j=1}^N\Big(\fs_{ij}\, E_{ij}\otimes E_{ji} +\fs_{ij}\,E_{ij}\otimes E_{ij} \Big)\,.
\end{split}
\end{equation}
In equations \eqref{eq:PB}, we have used the auxiliary space description: for any $N\times N$ matrix $M$, we define 
$M_1=M\otimes \II_N$ and $M_2=\II_N\otimes M= \Pi\, M_1\,\Pi$. Similarly, for any matrix $M_{12}\in \text{End}(\CC^n)\otimes \text{End}(\CC^n)$, we define 
$M_{21}=\Pi\,M_{12}\,\Pi$. In the following, we will call the algebra generated by $A$, $T$ and $S$ the \textbf{peakon algebra}.

The Poisson structure for the original Camassa-Holm model is recovered by setting $\nu=-1$. In that case, $T+A$ and $S$ form a Poisson subalgebra which yield the Camassa-Holm model.

\subsection{Spectral parameter presentation \label{sec:mat-r}}
It has been shown in \cite{AFR23} that the  peakon algebra can be presented through an $r$-matrix. 
In the present paper, it will be defined as
\begin{equation}\label{eq:PBspec}
\{\cL_1(z_1)\,,\, \cL_2(z_2)\}_{(1)} = [\fr_{12}(z_1, z_2)\,,\, \cL_1(z_1)] -[\fr_{21}(z_2, z_1)\,,\, \cL_2(z_2)]\,,
\end{equation}
where the Lax matrix $\cL(z)$ reads
\begin{equation}\label{eq:Thalved}
\begin{split}
\cL(z)&= z\,T^{h}+z^{\sigma}\,(T^{h})^t+S\,,\qquad
z^\sigma = \sigma(z)=-\frac{\frac{\gamma}2 z+1}{z+\frac{\gamma}2}\,\\
T^{h}=&\frac12\sum_{i,j=1}^N \sqrt{p_ip_j}\, e^{\frac{\nu}2(q_i-q_j)}\,E_{ij} \,,\qquad
S= \sum_{i,j=1}^N \sqrt{p_ip_j} \,\fs_{ij}\,\sinh \frac{\nu}{2}(q_i-q_j) \, E_{ij} \,,
\end{split}
\end{equation}
while the $r$-matrix has the form
\begin{equation}\label{rmat-z}
\begin{split}
 \fr_{12}(z_1, z_2)=& \frac{\nu}4 \Big( \frac{z_1z_2-1}{z_1-z_2}\,\Pi 
 +\frac{z_1z_2^{\sigma}-1}{z_1-z_2^{\sigma}}\,\Pi^t -\Gamma_{12} \Big)\,.
\end{split}
\end{equation}
The notation $T^h$ stands for ''halved $T$ matrix'' since $T$ in \eqref{eq:T} decomposes as $T^h+(T^h)^t$.

The connexion between \eqref{rmat-z} and the $r$-matrix defined in \cite{AFR23} is obtained through a redefinition of the  parameters as
\begin{equation}
\label{eq:z-x}
\begin{split}
z_j&= \frac{\frac{\gamma}2-\alpha-(\frac{\gamma}2+\alpha)\tanh(x_j/2)}{\tanh(x_j/2)-1}\,,\quad j=1,2\,,
\\
\alpha^2&=1-\frac{\gamma^2}4\qquad;\qquad\gamma=\frac{\lambda^2-\rho^2-1}{\rho}\,,
\end{split}
\end{equation}
 where $x_1$ and $x_2$ are the spectral parameters used in \cite{AFR23} and $\lambda$ and $\rho$ are the supplementary parameters of the Lax matrix introduced in \cite{AFR23}. Note that this transformation is not invertible when $\gamma=\pm2$: we treat this limiting case in Section \ref{sect:peak-strict}, and otherwise suppose that $\gamma\neq\pm2$.
 
 We characterize here $\gamma$ as a ``deformation'' parameter, allowing to generalize the strict peakons and remaining constant in $\cL(\gamma,z)$, and $z$ as a variable spectral parameter in \eqref{eq:PBspec}.

We remind that the $r$-matrix \eqref{rmat-z} obeys the classical (non skew-symmetric) Yang--Baxter equation
\begin{equation}\label{eq:cybe}
\big[ \fr_{12}(z_1, z_2)\,,\,\fr_{13}(z_1, z_3)  +\fr_{23}(z_2, z_3) \big] 
+ \big[ \fr_{32}(z_3, z_2)\,,\,\fr_{13}(z_1, z_3) \big]=0\,.
\end{equation}

\subsection{Degeneration to the original peakon model\label{sect:peak-strict}}
The linear Poisson structure of the original peakon model is given by
\begin{equation}\label{eq:PB-peakon}
\{L_1\,,\, L_2\}_{(1)} = -\frac{\mu}4\Big( [\Gamma_{12}\,,\, L_1] -[\Gamma_{21}\,,\, L_2] \Big)\,,
\end{equation}
where\footnote{Strictly speaking, we should set $\mu=-1$ to get the original peakon model. We keep $\mu$ to simplify the comparison with the spectral parameter approach.
The modulus $|\mu|$ can be set to 1 through a rescaling of the parameters $q_i$. }
\begin{equation}
\begin{split}
L&= \sum_{i, j=1}^N \sqrt{p_ip_j}\, e^{\frac{\mu}2|q_i-q_j|}\,E_{ij} \,.
\end{split}
\end{equation}
This structure can be obtained from the spectral parameter dependent one through the limit  $\gamma\to\pm 2$. In view of the expression \eqref{eq:Thalved}, $\gamma=\pm2$ implies that $\alpha=0$ and $z^\sigma=\mp 1$ for all $z$. Moreover,  
the expression \eqref{eq:z-x} at $\alpha=0$ leads to $z_j=\mp1$, $j=1,2$. Altogether, we get
\begin{equation}
\begin{split}
&\cL(z)\Big|_{\gamma=\pm2}=\mp\Big(T^{h}+(T^{h})^t\Big)+S= L\Big|_{\mu\to\mp\nu}.
\end{split}
\end{equation}

It remains to take the limit $\frac{\gamma}2\to\pm1$ in  $ \fr_{12}(z_1, z_2)$. We detail the case $\gamma=-2$, the second case being dealt analogously. We set $\frac{\gamma}2=-1+\eps^2$, $\eps\to0^+$.
Then from \eqref{eq:z-x}, we deduce 
\begin{equation}
\alpha=\eps\sqrt{2}+O(\eps^3)\ ,\quad 
z_j=1+\eps\sqrt{2}e^{x_j}+O(\eps^2)\ ,\quad z_j^\sigma=1-\eps\sqrt{2}e^{-x_j}+O(\eps^2)\,,\quad j=1,2\,.
\end{equation}
Plugging these expansions in $\fr_{12}(z_1, z_2)$, we get
\begin{equation}
\begin{split}
 \fr_{12}(z_1, z_2)=& \frac{\nu}4 \Big(\frac{e^{x_1}+e^{x_2}}{e^{x_1}-e^{x_2}}\, \Pi +\frac{e^{x_1+x_2}-1}{e^{x_1+x_2}+1}\,\Pi^t -\Gamma_{12} \Big)+O(\eps)\,,\\
 \fr_{21}(z_2, z_1)=& \frac{\nu}4 \Big(-\frac{e^{x_1}+e^{x_2}}{e^{x_1}-e^{x_2}}\, \Pi +\frac{e^{x_1+x_2}-1}{e^{x_1+x_2}+1}\,\Pi^t -\Gamma_{21} \Big)+O(\eps)\,.
\end{split}
\end{equation}
As noticed in  \cite{AFR22}, for any symmetric matrix $M$, we have 
\begin{equation}
 [\Pi \,,\, M_1] +[ \Pi \,,\, M_2]=0\,,\qquad  [\Pi^t\,,\, M_1] -[ \Pi^t \,,\, M_2]=0\,,
\end{equation}
so that, $L$ being symmetric, the relation \eqref{eq:PBspec} at leading order in $\eps$ leads to 
\begin{equation}
\begin{split}
&\{L_1\,,\, L_2\}_{(1)} = \frac{-\nu}4  \Big( [\Gamma_{12}\,,\, L_1] -  [\Gamma_{21}\,,\, L_2] \Big)\,,
\end{split}
\end{equation}
and we recover the relation \eqref{eq:PB-peakon}, with $\mu=\nu$.

Then, the Yang--Baxter equation \eqref{eq:cybe} for $ \fr_{12}(z_1, z_2)$ at leading order in $\eps$ leads to the same Yang--Baxter equation for the matrix
\begin{equation}\label{eq:newr}
 \mr_{12}(y_1, y_2)= \frac{\nu}4 \Big(\frac{y_1+y_2}{y_1-y_2}\, \Pi +\frac{y_1y_2-1}{y_1y_2+1}\,\Pi^t  -\Gamma_{12}\Big)\,,\quad
 \text{with $y=e^x$.}
\end{equation}
Extracting from this Yang--Baxter relation the quadratic terms in $\Gamma$, we get 
\begin{equation}\label{eq:cybe2}
\big[ \Gamma_{12}\,,\,\Gamma_{13}  +\Gamma_{23} \big] + \big[ \Gamma_{32}\,,\,\Gamma_{13} \big]=\textsc{s}_{123}(y_1,y_2,y_3)\,,
\end{equation}
where $\textsc{s}_{123}(y_1,y_2,y_3)$ gathers all the other terms.
It can be shown that $\textsc{s}_{123}(y_1,y_2,y_3)$ is indeed independent from $y_1,y_2$ and $y_3$ and takes the form
\begin{equation}\label{eq:2nd-membre}
\begin{split}
&\textsc{s}_{123}(y_1,y_2,y_3) =\textsc{c}_{123} + \textsc{c}_{123}^{t_2}+ \textsc{c}_{123}^{t_3}- \textsc{c}_{123}^{t_1} \,,\\
&\textsc{c}_{123}=\sum_{i,j,k=1}^N \Big(E_{ij}\otimes E_{jk}\otimes E_{ki}-E_{ji}\otimes E_{kj}\otimes E_{ik}\Big)\,,
\end{split}
\end{equation}
where $t_k$ is the transposition in the space $k$ ($k=1,2,3$).
Gathering \eqref{eq:cybe2} and \eqref{eq:2nd-membre}, we obtain a modified Yang--Baxter equation for  $\Gamma_{12}$, as expected from  \cite{AFR22}. 

Remark that since the $r$-matrix \eqref{eq:newr} obeys the Yang--Baxter relation, one can seek for Lax matrices representing its 
related Poisson structure and build integrable models upon it. 

\subsection{Degeneration to the Ragnisco--Bruschi peakon model\label{sect:peak-RB-lin}}
The Lax matrix $\cL(z)$ \eqref{eq:Thalved} is written in terms of the matrices $T,S,A$, see eqs \eqref{eq:A}--\eqref{eq:S}, as
\begin{equation}\label{eq:LRB1}
\cL(z)= \frac{z+z^{\sigma}}{2}\,T+S+\frac{z-z^{\sigma}}{2}\,A\,.
\end{equation}
The original Lax matrix of the Ragnisco--Bruschi peakon model \cite{RB} is $L^{\rho} = T + \rho S$. It follows that the Ragnisco--Bruschi Lax matrix is obtained from $\cL(z)$, up to a constant factor $\rho$, for the values of the spectral parameter $z$ such that $z^{\sigma}=z$, in other words for the fixed points of the involution $\sigma$: $z_{\pm}=\frac12\big(-\gamma\pm\sqrt{\gamma^2-4}\big)$. Note that $z_+z_-=1$. \\
Hence one gets (with $\rho_{\pm}=z_{\pm}$)
\begin{equation}
L^{\rho_\pm} = z_{\pm} \cL(z_{\mp}) = T + z_{\pm}\, S \,.
\end{equation}
Since the matrix $\fr_{12}(z_1,z_2)$ is singular at $z_2=z_1$, the Poisson structure for the Ragnisco--Bruschi Lax matrix $\{L^{\rho_+}_{1},L^{\rho_+}_{2}\}$ (and similarly for $L^{\rho_-}$) cannot be obtained naively considering \eqref{eq:PBspec}. However, taking $z_1=z_+$ and $z_2=z_-$, the matrix $\fr_{12}(z_+,z_-)$ is regular and reduces to $-\frac{\nu}{4}\,\Gamma_{12}$ (thanks to the property $z_+z_-=1$ and the fact that $z_\pm$ are fixed points of the involution $\sigma$). In that case, one gets the linear Poisson structure \begin{equation}\label{eq:PBRB}
\{L^{\rho_+}_{1},L^{\rho_-}_{2}\}_{(1)} = -\frac{\nu}{4} \Big( [\Gamma_{12}\,,\, L^{\rho_+}_{1}] -[\Gamma_{21}\,,\, L^{\rho_-}_{2}] \Big) \,.
\end{equation}
Using the representation \eqref{eq:LRB1}, one gets the Ragnisco--Bruschi linear Poisson brackets for the variables $p_k$ and $q_{ij}$, which follow from the canonical Poisson brackets for $p_k$ and $q_{i}$.

\section{Halved algebra\label{sec:split_alg}}
The identification in \eqref{eq:Thalved} of two separate components of $T$ related by a morphism can be extended to the full Lax matrix $L$. It yields a remarkable Poisson algebra structure (the halved algebra) for the ``half'' Lax matrix with a separated $r$-matrix, that obeys the classical Yang--Baxter equation. Its quadratic extension is easily deduced.
\subsection{Poisson brackets of the halved algebra}
In the same way as $T$, see \eqref{eq:Thalved}, one can split $S$ as $S^h+(S^h)^t$, with 
 \begin{equation}\label{eq:split-mat}
\begin{split}
S^{h}&= \frac12\sum_{i, j=1}^N \sqrt{p_ip_j}\, \fs_{ij}\,e^{\frac{\nu}2(q_i-q_j)}\,E_{ij} \,.
\end{split}
 \end{equation}
It leads to
 \begin{equation}\label{eq:Lsplity}
\begin{split}
\cL(z)=\ell(z)+\ell(z^\sigma)^t\,\quad\text{ with }\quad \ell(z) &= z\,T^{h}+S^{h}\,,
\end{split}
 \end{equation}
or in component
 \begin{equation}\label{eq:Lsplity2}
\begin{split}
 \ell(z) &= \sum_{ij=1}^N \ell_{ij}(z)\,E_{ij}\quad\text{ with }\quad 
 \ell_{ij}(z)=\frac12\sqrt{p_ip_j}\, \big( z+\fs_{ij}\big)\,e^{\frac{\nu}2(q_i-q_j)}\,.
\end{split}
 \end{equation}

Similarly, it is clear that the $r$-matrix \eqref{rmat-z} can be split as 
\begin{equation}\label{eq:Rsplity}
\begin{split}
\fr(z_1, z_2)&=r(z_1,z_2)+r(z_1,z^\sigma_2)^{t_2}  \,,\\
r(z_1,z_2)&= \frac{\nu}4 \Big(\frac{z_1z_2-1}{z_1-z_2}\,\Pi -P\Big) \,,\\
\end{split}
 \end{equation}
 with $P$ given by:
\begin{equation}\label{eq:Psigned}
\begin{split}
P&=\sum_{i,j=1}^N \fs_{ij}\,E_{ij}\otimes E_{ji}  \,,
\end{split}
 \end{equation}
 so that $\Gamma_{12}=P+P^t$.
 $P$ coincides with  the classical $r$-matrix for the Toda model in the first chamber $q_i>q_j$ for $i>j$.
 
 Direct calculations show that the Poisson brackets of $T^{h}$ and $S^{h}$  take exactly the same form as the one given in Section \ref{sec:PB}, with the replacements
\begin{equation}
\begin{split}
&T\ \to\ T^{h}\ ;\quad S\ \to\ S^{h}\ ;\quad A\ \to\ T^{h}\,,\\
&\Pi\ \to\ \Pi\ ;\quad \Pi^t\ \to\ 0\ ;\quad \Gamma_{12}\ \to\ P_{12}\ ;\quad 
\Gamma_{21}\ \to\ P_{21}=-P_{12}\,,
\end{split}
\end{equation}
By direct computation one shows that these Poisson brackets are encoded in the relation
\begin{equation}\label{eq:PB0ell}
\big\{\ell_1(z_1)\,,\,\ell_2(z_2)\big\}_{(1)} = \big[r_{12}(z_1,z_2) \,,\, \ell_1(z_1)\big] - \big[r_{21}(z_2,z_1) \,,\, \ell_2(z_2)\big]\,.
\end{equation}
To prove this relation, we use the following properties
\begin{eqnarray}
\big[\Pi,M_1-M_2\big]&=& \frac{2}{k(z_1)-k(z_2)}\big[\Pi\,,\, k(z_1)\,M_1+ k(z_2)\,M_2\big]\,,\label{prop:Pif}\\
\big[\Pi,M_1+M_2\big]&=& 0\,,\label{prop:PiMM}
\end{eqnarray}
which are valid for any matrix $M$ (not depending on $z$), and any scalar function $k(z)$.

Explicitly, the Poisson brackets take the form
\begin{equation}
\begin{split}
\big\{\ell_{ij}(z_1)\,,\,\ell_{kl}(z_2)\big\}_{(1)} =&\quad \frac{\nu}4\delta_{il}\Big[\big(\frac{z_1z_2-1}{z_1-z_2}+\fs_{kl}\big)\,\ell_{kj}(z_1) - \big(\frac{z_1z_2-1}{z_1-z_2}-\fs_{ij}\big)\,\ell_{kj}(z_2)
\Big]\\
& -\frac{\nu}4\delta_{kj}\Big[\big(\frac{z_1z_2-1}{z_1-z_2}+\fs_{kl}\big)\,\ell_{il}(z_1) - \big(\frac{z_1z_2-1}{z_1-z_2}-\fs_{ij}\big)\,\ell_{il}(z_2)
\Big]
\end{split}
\end{equation}
which is compatible with the canonical Poisson structure \eqref{eq:PBcano}
and the expression \eqref{eq:Lsplity2} for $\ell(z)$.

The dynamics associated to this linear Poisson structure of the halved algebra is presented in appendix \ref{app:hamil}.

\subsection{Reconstruction of the peakon algebra from the halved algebra}
Starting from \eqref{eq:PB0ell}, by suitably applying the transposition in space 1 and/or 2,  and the involution $\sigma$ on the variables $z_1$ and/or $z_2$, 
 we get three additional relations
\begin{equation}\label{eq:PBellsigma}
\begin{split}
\big\{\ell_1(z_1)\,,\,\ell_2(z^\sigma_2)^{t_2}\big\}_{(1)} &= \big[r_{12}(z_1,z^\sigma_2)^{t_2} \,,\, \ell_1(z_1)\big] + \big[r_{21}(z^\sigma_2,z_1)^{t_2} \,,\, \ell_2(z^\sigma_2)^{t_2}\big]\,,\\
\big\{\ell_1(z^\sigma_1)^{t_1}\,,\,\ell_2(z_2)\big\}_{(1)} &= -\big[r_{12}(z^\sigma_1,z_2)^{t_1} \,,\, \ell_1(z^\sigma_1)^{t_1}\big] - \big[r_{21}(z_2,z^\sigma_1)^{t_1} \,,\, \ell_2(z_2)\big]\,,\\
\big\{\ell_1(z^\sigma_1)^{t_1}\,,\,\ell_2(z^\sigma_2)^{t_2}\big\}_{(1)} &= -\big[r_{12}(z^\sigma_1,z^\sigma_2)^{t_1t_2} \,,\, \ell_1(z^\sigma_1)^{t_1}\big] + \big[r_{21}(z^\sigma_2,z^\sigma_1)^{t_1t_2} \,,\, \ell_2(z^\sigma_2)^{t_2}\big]\,.
\end{split}
\end{equation}
From the explicit expression of $z^\sigma$, a simple calculation shows that 
\begin{equation}
\frac{z_1z_2-1}{z_1-z_2}
= -\frac{z_1^{\sigma}z_2^{\sigma}-1}{z_1^{\sigma}-z_2^{\sigma}}\,,\qquad
\frac{z_1^{\sigma}z_2-1}{z_1^{\sigma}-z_2}=-\frac{z_1z_2^{\sigma}-1}{z_1-z_2^{\sigma}}\,,
\end{equation}
which in turn implies
\begin{equation}\label{eq:prop_r_sym}
r_{12}(z^\sigma_1,z^\sigma_2)^{t_1t_2}=-r_{12}(z_1,z_2)
 \mb{and} r_{12}(z^\sigma_1,z_2)^{t_1}=-r_{12}(z_1,z^\sigma_2)^{t_2}\,.
\end{equation}

Then, adding the four relations \eqref{eq:PB0ell} and \eqref{eq:PBellsigma} we recover the relation \eqref{eq:PBspec} for the peakon algebra generated by $T$, $S$ and $A$ where the expression of $\ell(z)+\ell(z^\sigma)^{t}$ consistently provides
\begin{equation}\label{eq:TS-halved}
\begin{split}
T&= T^{h} + (T^{h})^t\ ;\quad A= T^{h} - (T^{h})^t\ ;\quad S= S^{h} + (S^{h})^t\, . 
\end{split}
\end{equation}
As can be seen from relations \eqref{eq:Lsplity} and \eqref{eq:Rsplity}, the peakon algebra appears as a classical twist of the halved algebra, in the spirit of a classical version of twisted Yangians \cite{MNO} and coideal subalgebras \cite{coideal}.

\subsection{Yang--Baxter equation}
A direct calculation shows that the $r$-matrix \eqref{eq:Rsplity} obeys the classical Yang--Baxter equation:
\begin{equation}
\big[ r_{12}(z_1,z_2)\,,\, r_{13}(z_1,z_3)+r_{23}(z_2,z_3)\big]
+\big[ r_{32}(z_3,z_2)\,,\, r_{13}(z_1,z_3)\big] =0\,,
\end{equation}
together with the antisymmetry relation
\begin{equation}
r_{12}(z_1,z_2)=- r_{21}(z_2,z_1)\,.
\end{equation}

Using antisymmetry, we can rewrite the Yang--Baxter equation as
\begin{equation}
\begin{split}
&\big[r_{13}(z_1,z_3)\,,\,  r_{23}(z_2,z_3)\big]=
\big[ r_{12}(z_1,z_2)\,,\, -r_{13}(z_1,z_3)\big]-
\big[ r_{21}(z_2,z_1)\,,\, -r_{23}(z_2,z_3)\big]\,.
\end{split}
\end{equation}
This shows that one can represent the halved algebra using the $r$-matrix and the commutator:
\begin{equation}
\begin{split}
&\ell_1(z_1)\ \to\ -r_{13}(z_1,z_3)\,,\\
&\big\{\ell_1(z_1)\,,\, \ell_2(z_2)\big\}_{(1)}\ \to\ \big[r_{13}(z_1,z_3)\,,\,  r_{23}(z_2,z_3)\big]\,,
\end{split}
\end{equation}
where $z_3$ plays the role of an inhomogeneity parameter.

\subsection{Quadratic Poisson brackets for the halved algebra\label{sec:split_quad}}
Starting from the $r$-matrix 
 \begin{equation}\label{eq:rmat-split}
 r(z_1,z_2)= \frac{\nu}4\Big( \frac{z_1z_2-1}{z_1-z_2}\,\Pi - P\Big)= \frac{\nu}4\,\sum_{j,k=1}^N \Big( \frac{z_1z_2-1}{z_1-z_2} - \fs_{jk}\Big)E_{jk}\otimes E_{kj}\,,
\end{equation}
we can define a second Poisson structure
\begin{equation}
\big\{\ell_1(z_1)\,,\,\ell_2(z_2)\big\}_{(2)} = \big[r_{12}(z_1,z_2)\,,\,\ell_1(z_1)\,\ell_2(z_2)\big]\,,
\end{equation}
or equivalently
\begin{equation}\label{eq:PB2ell}
\begin{split}
\big\{\ell_{ij}(z_1)\,,\,\ell_{kl}(z_2)\big\}_{(2)}  =\ & \frac{\nu}4\,\Big( \frac{z_1z_2-1}{z_1-z_2} - \fs_{ik}\Big)\ell_{kj}(z_1) \ell_{il}(z_2)\\
&-\frac{\nu}4\,\Big( \frac{z_1z_2-1}{z_1-z_2} - \fs_{lj}\Big)\ell_{kj}(z_2) \ell_{il}(z_1)\,.
\end{split}
\end{equation}
Since $ r(z_1,z_2)$ is antisymmetric and obeys the Yang--Baxter equation, the Poisson structure is well-defined.

The Hamiltonians corresponding to this second quadratic Poisson structure are presented in appendix \ref{app:Ham2-halved}.

\section{Quadratic Poisson brackets for the peakon model\label{sec:quad-peakon}}

\subsection{Quadratic structure induced by the halving of the algebra}
We can now reconstruct a quadratic Poisson structure for the generalized peakon model from the results of Section \ref{sec:split_alg} on the halved algebra.
We  introduce
\begin{equation}
L(z)=\ell(z)\,\ell(z^{\sigma})^t\,.
\end{equation}
Using the relations \eqref{eq:prop_r_sym} and the fact that $\sigma$ is an antimorphism, 
it is easy to show that 
\begin{equation}\label{eq:rll-quad}
\begin{split}
\big\{L_1(z_1)\,,\,L_2(z_2)\big\}_{(2)} =\ & \big[r_{12}(z_1,z_2)\,,\,L_1(z_1)\,L_2(z_2)\big]\\
&+L_1(z_1)\,r_{12}(z^\sigma_1,z_2)^{t_1}\,L_2(z_2) -L_2(z_2)\,r_{12}(z^\sigma_1,z_2)^{t_1}\,L_1(z_1)
\,,
\end{split}
\end{equation}
where $r_{12}(z_1,z_2)$ is given in \eqref{eq:rmat-split}.
Equivalently
\begin{equation}
\label{eq:PB-quad}
\begin{split}
\big\{L_{ij}(z_1)\,,\,L_{kl}(z_2)\big\}_{(2)} =\ & \frac{\nu}4\,\Big( \frac{z_1z_2-1}{z_1-z_2} - \fs_{ik}\Big)L_{kj}(z_1) L_{il}(z_2)\\
&-\frac{\nu}4\,\Big( \frac{z_1z_2-1}{z_1-z_2} - \fs_{lj}\Big)L_{kj}(z_2) L_{il}(z_1)\\
&+\frac{\nu}4\Big( \frac{z_1^\sigma z_2-1}{z_1^\sigma-z_2} - \fs_{jk}\Big)L_{ik}(z_1) L_{jl}(z_2)\\
&-\frac{\nu}4\Big( \frac{z_1^\sigma z_2-1}{z_1^\sigma-z_2} - \fs_{li}\Big)L_{ki}(z_2) L_{lj}(z_1)\,,
\end{split}
\end{equation}
where $z^{\sigma}$ is given in \eqref{eq:Thalved}.

\medskip
Taking the form \eqref{eq:Lsplity} for $\ell(z)$ and setting $\bar p=\{p_1,p_2,...,p_N\}$ and $\bar q=\{q_1,q_2,...,q_N\}$, we get a representation
\begin{align}
L(z) &= \sum_{i,j=1}^N \Lambda_{ij}(z,\bar p,\bar q)\,\sqrt{p_ip_j}\,e^{\frac{\nu}2(q_i+q_j)}\,E_{ij}\,,\\
\Lambda_{ij}(z,\bar p,\bar q) &=\sum_{k=1}^N (z+\fs_{ik})(z^\sigma+\fs_{jk})\,p_k\,e^{-{\nu}q_k}\,,
\label{eq:Lambda}
\end{align}
where $q_i$ and $p_j$ obey the Poisson brackets of the halved algebra, as given in \eqref{eq:PBell}.

Although obeying the relation \eqref{eq:PB-quad}, this representation is not local, in the sense that $L_{ij}(z)$ depends 
not only on $p_i$, $p_j$, $q_{i}$, and $q_{j}$, but rather on all $p_k$ and $q_{k}$, $k=1,...,N$, see expression \eqref{eq:Lambda}. Our aim is now to build a 
representation of \eqref{eq:PB-quad} by a Lax matrix obeying this locality property. This Lax matrix can then be characterized as a ``peakon'' representation for the quadratic structure.

\subsection{Constructing a peakon representation\label{sec:PB-quad}}
We first remark that if one takes a constant Lax matrix $L$ such that $L^t=\pm L$, then the properties of $\Pi$ and $\Pi^t$ allow to rewrite \eqref{eq:rll-quad} as
\begin{equation}\label{eq:rll-quad-cst}
\begin{split}
\big\{L_1\,,\,L_2\big\}_{(2)} =\ & -\frac{\nu}4\Big(\big[P\,,\,L_1\,L_2\big]+L_1\,P^{t}\,L_2 -L_2\,P^t\,L_1\Big)
\,,\\
\big\{L_{ij}\,,\,L_{kl}\big\}_{(2)} =\ & \frac{\nu}4\,\Big((\fs_{lj}-\fs_{ik})L_{kj} L_{il}  +(\fs_{li}-\fs_{jk})L_{ik} L_{jl}\Big)
\end{split}
\end{equation}
and we get a realization from the quadratic Camassa--Holm peakon structure for the Lax matrix element $L_{ij}=\sqrt{p_ip_j}e^{-\frac{\nu}2|q_i-q_j|}$ \cite{AFR22, dGHH}:
\begin{equation}\label{eq:PBCH} 
\begin{aligned}
\{ {p}_i,{p}_j \}_{(2)} &= -\fs_{ij}\,\nu\,{p}_i {p}_j e^{-\nu|{q}_i-{q}_j| } \,, \\
\{ {q}_i,{p}_j \}_{(2)} &= -{p}_j e^{-\nu|{q}_i-{q}_j|} \,,\qquad 
\\
\{ {q}_i,{q}_j \}_{(2)} &= -\frac1{\nu}\,\fs_{ij} \big( 1-e^{-\nu|{q}_i-{q}_j| }\big) \,.
\end{aligned}
\end{equation}
Precisely the Poisson structure \eqref{eq:rll-quad-cst} yields a Poisson bracket for the relevant variables $p_i, q_ {ij}$ of $L$.  This Poisson bracket corresponds to the Camassa--Holm second Poisson bracket \eqref{eq:PBCH}. 
Note that $L$ depends only on the relative peakon
positions $q_i - q_j \equiv q_{ij}$ but in this case the complete set of variables $\{ p_i, q_ {j} \}$ carries a genuine (associative) Poisson structure. This will not be the case for the generalized peakons.

Now we consider the Lax matrix \eqref{eq:Thalved} with entries 
\begin{equation}\label{eq:L-linear}
\begin{split}
\cL_{ij}(z)=&\sqrt{p_ip_j}\Big(h^+_{ij}(z)\,e^{\frac{\nu}2|q_{ij}|}+h^-_{ij}(z)\,e^{-\frac{\nu}2|q_{ij}|}\Big)\,,\\ 
h^+_{ij}(z)=&\frac{1}{2}\Big(1+\frac{z+z^{\sigma}}2+\fs_{ij}\frac{z-z^{\sigma}}2\Big)
= \frac{1}{2}\Big( 1+\frac{z^2-1}{2z+\gamma}+\fs_{ij}\frac{z^2+\gamma z+1}{2z+\gamma} \Big) \,, \\[.4ex]
h^-_{ij}(z)=&\frac{1}{2}\Big(-1+\frac{z+z^{\sigma}}2-\fs_{ij}\frac{z-z^{\sigma}}2\Big)
= \frac{1}{2}\Big( -1+\frac{z^2-1}{2z+\gamma}-\fs_{ij}\frac{z^2+\gamma z+1}{2z+\gamma} \Big) \,,
\end{split}
\end{equation}
which is defined on the phase space spanned by $p_i$, $i=1,...,N$ and $q_{ij}=q_i-q_j$, $1\leq i\neq j\leq N$. We will see that the resulting Poisson structure is compatible with the constraints $q_{ji}=-q_{ij}$ and $q_{ik}=q_{ij}+q_{jk}$. However, we do not look for a Poisson structure including the center of mass $\sum_{i=1}^Nq_i$. As such, the Poisson structure is not defined of the phase space $\cP$ 
generated by $p_i$ and $q_j$, but rather on the coset $\cP/\cT$, where $\cT$ is the set of translations $q_j\to q_j+q_0$, $\forall j$.

Plugging it in the relations \eqref{eq:PB-quad}, and following the general procedure described in the Appendix \ref{appB}, one obtains 
\begin{equation}
\begin{split}
\big\{ p_i\,,\,p_k\big\}_{(2)} =\ & \frac{\nu}{4}\,\fs_{ik}\,{p_ip_k}\big( \cQ_{ik}-4\big)\,,
\\
\big\{ q_{ij}\,,\,p_k\big\}_{(2)} =\ & -\frac{1}{4}{p_k}\Big( \cS_{ik}-\cS_{jk}\Big)\,,\\
\big\{ q_{ij}\,,\,q_{kl}\big\}_{(2)} =\ & -\frac{1}{4\nu}\Big(\fs_{ik}\cQ_{ik}-\fs_{jk}\cQ_{jk}-\fs_{il}\cQ_{il}+\fs_{jl}\cQ_{jl}\Big)\,,
\end{split}
\label{eq:qijpk}
\end{equation}
where we have introduced 
\begin{equation}\label{eq:SQ}
\begin{split}
\cQ_{ik} =\ & (\gamma-2)e^{{\nu}|q_{ik}|} - (\gamma+2)e^{-{\nu}|q_{ik}|} + 4 
\,, \\
\cS_{ik} =\ & (\gamma-2)e^{{\nu}|q_{ik}|} + (\gamma+2)e^{-{\nu}|q_{ik}|} 
\,.
\end{split}
\end{equation}
Relations \eqref{eq:qijpk} obey the Jacobi identities since they derive analytically (through the equations of Appendix A) from the $r$-matrix Poisson structure of the Lax matrix. 

\begin{rmk}\label{rmk:cyclic}
The relations \eqref{eq:qijpk} involve $2N-1$ variables $p_i$ and $q_{i,i+1}$. 
Since we have a Poisson bracket on this $(2N-1)$-dimensional phase space, it implies that there exists a cyclic variable, which Poisson-commutes with all variables. We will come back on this point in section \ref{sect:cycl}.
\end{rmk}

The  consistency of the Poisson structure \eqref{eq:qijpk} with respect to the relation $q_{ij}=q_i-q_j$ is guaranteed by the form of the r.h.s. of \eqref{eq:qijpk}.  For instance, the Poisson bracket $\{q_{ij}\,,\,p_k\}_{(2)}$ should be consistently obtained from the Poisson brackets $\{q_{i\ell}\,,\,p_k\}_{(2)}+\{q_{\ell j}\,,\,p_k\}_{(2)}$, which is ensured by the form $\cS_{ik}-\cS_{jk}$ occurring in the r.h.s. of \eqref{eq:qijpk}. The same is true for the Poisson bracket $\{q_{ij}\,,\,q_{k\ell}\}_{(2)}$.

Note that the Poisson brackets $\{q_{ij}, p_k\}_{(2)}$ and $\{q_{ij}, q_{kl}\}_{(2)}$ can be expressed as linear combinations of the bilinears $(q_i,p_k)$ and $(q_i,q_k)$, which can be generalized to the form:  
\begin{equation}\label{eq:qipk}
\begin{split}
(q_{i}\,,\,p_k) = &-\frac{1}{4}{p_k}\Big( (\gamma-2)e^{{\nu}|q_i-q_k|} + (\gamma+2)e^{-{\nu}|q_i-q_k|} +\lambda_k\Big)\,,\\
(q_i\,,\,q_k) =\,& -\frac{1}{4\nu}\fs_{ik}\Big( (\gamma-2)e^{{\nu}|q_i-q_k|} - (\gamma+2)e^{-{\nu}|q_i-q_k|} + J_{ik}\Big) +\alpha_i-\alpha_k\,,
\end{split}
\end{equation}
where
\begin{equation}
J_{ik} = 4(1 - |k-i|) + \sum_{l=\min(i,k)}^{\max(i,k)-1}J_{l,l+1} \,.
\end{equation}
However \eqref{eq:qipk} cannot be promoted as genuine Poisson brackets for the local peakon variables $p_i, q_j$. 
Indeed, we found that when $\gamma\neq\pm2$, there are no values for $\lambda_k$, $J_{ik}$ and $\alpha_i$ that fulfill the Jacobi identities for the variables $p_i$ and $q_i$, although the relations \eqref{eq:qijpk} fulfill the Jacobi identities for the variables $p_i$ and $q_{ij}$. In other words, when $\gamma\neq\pm2$ the quadratic structure does not allow for a dynamical center of mass, only the relative positions $q_{ij}$ are compatible with this structure.

\begin{rmk}\label{rmk:quadCH}
\textbf{Reduction to the original quadratic structure of Camassa--Holm peakons.} For $\gamma=\pm2$, we recover the quadratic structure of Camassa--Holm peakons \eqref{eq:PBCH} for the values
$\lambda_k=0$, $\alpha_k=0$ and $J_{l,l+1}=4$. In that case, the Jacobi identities for $p_i$ and $q_i$ are fulfilled.
\end{rmk}

\begin{rmk}
It is worth noticing that the Poisson structure \eqref{eq:qijpk} takes a form similar to the generalized Poisson brackets for local peakon variables $q_i,p_j$ investigated in the Appendix of \cite{HH}:
\begin{equation}\begin{split}
\{q_{i},q_{j}\}_{(2)} = G(q_{ij})\,,\quad
\{q_{i},p_j\}_{(2)}= p_j\,G'(q_{ij})\,,\quad
\{p_i,p_j\}_{(2)} = -p_ip_j\,G''(q_{ij})\,.
 \end{split}
 \end{equation}
Indeed introducing the function $G(x)= \sgn(x)\big( (\gamma-2)e^{{\nu}|x|} - (\gamma+2)e^{-{\nu}|x|} + 4 \big)$ such that
the Poisson brackets $\{q_{ij},q_{kl}\}_{(2)}$ are controlled by $G (q_{ab})  \equiv \fs_{ab}\,\cQ_{ab}$, one can check that the brackets $\{q_{ij},p_k\}_{(2)}$ are controlled by $G'(q_{ab})$ and the brackets $\{p_i,p_k\}_{(2)}$ by  $-G''(q_{ab})$. However in \cite{HH} the bracket structure under investigation is defined for the variables $p_i,q_j$ which we have seen to be non consistent in our case.
This explains why our function $G$ is not in  the classification presented in \cite{HH}. 
Remark that $G'(x)$ and $G''(x)$ are well-defined in 0 because $G(x)=\sgn(x)H(x)$ with $H(0)=0$ and the convention $\sgn(0)=0$.
\end{rmk}

\subsection{A quadratic structure for the Ragnisco--Bruschi peakons\label{sect:peak-RB-quad}}
As explained in section \ref{sect:peak-RB-lin}, the Ragnisco--Bruschi Lax matrix is given by taking for the spectral parameter $z$ in $\cL(z)$ the values of the fixed points of the involution $\sigma$. Since the $r$-matrices that appear in the quadratic structure \eqref{eq:PB-quad} have poles in $z_2=z_1$ and $z_2=z_1^{\sigma}$, it is again necessary to consider the quadratic structure \eqref{eq:PB-quad} in the two fixed points $z_1=z_+$ and $z_2=z_-$. It follows that the $z$-dependence in \eqref{eq:PB-quad} disappears and the Ragnisco--Bruschi Lax matrices satisfy the following quadratic Poisson structure:
\begin{equation}
\big\{L^{\rho_+}_1\,,\,L^{\rho_-}_2\big\}_{(2)} = -\frac{\nu}4\Big(\big[P\,,\,L^{\rho_+}_1\,L^{\rho_-}_2\big]+L^{\rho_+}_1\,P^{t}\,L^{\rho_-}_2 -L^{\rho_-}_2\,P^t\,L^{\rho_+}_1\Big)\,,
\end{equation}
or equivalently
\begin{equation}
\label{eq:PB-RB-quad}
\begin{split}
\{L^{\rho_+}_{ij},L^{\rho_-}_{kl}\}_{(2)} = -\frac{\nu}4\,&\Big( \fs_{ik}\,L^{\rho_+}_{kj}\,L^{\rho_-}_{il} + \fs_{jl}\,L^{\rho_-}_{kj}\,L^{\rho_+}_{il} 
+ \fs_{jk}\,L^{\rho_+}_{ik}\,L^{\rho_-}_{jl} + \fs_{il}\,L^{\rho_-}_{ki}\,L^{\rho_+}_{lj} \Big) \,.
\end{split}
\end{equation}
It leads to a quadratic Poisson structure for the variables $p_k$ and $q_{ij}$. To the best of our knowledge, this quadratic Poisson structure was not known for the Ragnisco--Bruschi peakons.

Let us emphasize that the structure \eqref{eq:PB-RB-quad} is different from the one in \eqref{eq:rll-quad-cst}. Indeed, in the Camassa--Holm case, the same Lax matrix appears in the two auxiliary spaces. On the contrary, in the Ragnisco--Bruschi case, one needs to consider two different Lax matrices, one corresponding to $\rho_{+}=z_+$ and one corresponding to $\rho_{-}=z_-$, in order to define a quadratic Poisson structure. Hence the Poisson brackets for the variables $p_k, q_ {ij}$ are not given by the original peakon Poisson brackets \eqref{eq:PBCH}, although the matrices $L^{\rho_\pm}$ are constant matrices such that $(L^{\rho_\pm})^t=L^{\rho_\pm}$ as in the reduction done in section \ref{sect:peak-strict}. They instead obey the general formulas \eqref{eq:qijpk}. The need for two different Lax matrices to define the quadratic structure explains why such a structure has not been proposed before. In this regard, the spectral parameter approach presented in this paper is more natural, since the quadratic structure is obtained through a single Lax matrix.

When $\gamma=\pm2$, one recovers the quadratic structure of the Camassa--Holm peakons \eqref{eq:PBCH} as expected.

\section{Hamiltonian dynamics associated to the quadratic structure\label{sec:hamil}}
\subsection{Hamiltonian structure and time evolution\label{sec:ham-quad}}
From the Lax matrix $\cL(z)$, see \eqref{eq:L-linear}, one introduces the transfer matrices
\begin{equation}
\tau_m(z)=\tr\big(\cL(z)^{m}\big)=\sum_{i_1,i_2,...,i_{m-1}=1}^N \cL_{i_1,i_2}(z)\,\cL_{i_2,i_3}(z)\,\cdots\cL_{i_{m-1},i_1}(z)\,.
\end{equation}
From the expression of $\cL(z)$, it is clear that $\tau_m(z)$ is an homogeneous, symmetric function of $p_1,...,p_N$ of degree $m$. 

The transfer matrices obey
\begin{equation}
\big\{\tau_{m_1}(z_1)\,,\,\tau_{m_2}(z_2)\big\}_{(2)} =0 \,.
\end{equation}
Indeed, one has
\begin{equation}
\begin{split}
& \big\{\tau_{m_1}(z_1)\,,\,\tau_{m_2}(z_2)\big\}_{(2)} 
=\ m_1m_2\, \tr_{12}\Big( [r_{12}(z_1,z_2)\,,\, \cL_1(z_1)^{m_1}\,\cL_2(z_2)^{m_2}]\\
&\qquad + \cL_1(z_1)^{m_1}\,r_{12}(z_1^{\sigma},z_2)^{t_1}\,\cL_2(z_2)^{m_2}
 -\cL_2(z_2)^{m_2}\,r_{12}(z_1^{\sigma},z_2)^{t_1}\,\cL_1(z_1)^{m_1}\Big)\\
&\quad=\ 0\,.
\end{split}
\end{equation}
The last equality is obtained using the cyclicity of the trace and $[\cL_1(z_1),\cL_2(z_2)]=0$. 

\subsection{Examples}
Computing directly and explicitly $\tau_m(z)$ for general $m$ seems a bit difficult to achieve. In what follows, we provide explicit expressions for the first transfer matrices, the associated Hamiltonians and corresponding time evolutions.
We introduce the functions 
\begin{equation}\label{eq:Lambda-xi}
\Lambda(z)=\dfrac{z+z^{\sigma}}{2}=\dfrac{z^2-1}{2z+\gamma}\,,\qquad \xi(z)=\dfrac{z-z^{\sigma}}{2}=\dfrac{z^2+\gamma z+1}{2z+\gamma}\,.
\end{equation}

The first transfer matrix takes the form
\begin{equation}
\tau_1(z) = \sum_{i=1}^N\cL_{ii}(z)=\Lambda(z)\,\fh^{(1)}\,,
\end{equation}
where the Hamiltonian $\fh^{(1)}$ is given by
\begin{equation}
\fh^{(1)} = \sum_{i=1}^N p_i \,.
\end{equation}
The corresponding equations of motion read
\begin{align}
\dot p_k &= \{\fh^{(1)}\,,\,p_k\}_{(2)} = \frac{\nu}{4}\,p_k \sum_{i=1}^N p_i\,\fs_{ik}(\cQ_{ik}-4) \,, \\
\dot q_{kl} &= \{\fh^{(1)}\,,\,q_{kl}\}_{(2)} = \frac14\,\sum_{i=1}^N p_i (\cS_{ik}-\cS_{il}) \,,
\end{align}
where $\cQ_{ik}$ and $\cS_{ik}$ are given in \eqref{eq:SQ}.

The next transfer matrix reads
\begin{equation}
\tau_2(z) = \sum_{i,j=1}^N\cL_{ij}(z)\cL_{ji}(z) = \Lambda(z)\,\fh^{(2)} + \Lambda(z)^2 \big(\fh^{(1)}\big)^2\,,
\end{equation}
where the Hamiltonian $\fh^{(2)}$ is given by
\begin{equation}
\fh^{(2)} = \sum_{i,j=1}^N \sfrac{1}{4}\,p_i\,p_j \big(2\gamma-\cS_{ij}\big)=\sum_{i,j=1}^N \sfrac{1}{2}\,p_i\,p_j\big(2\sinh (\nu|q_{ij}|)-\gamma\cosh (\nu|q_{ij}|)+\gamma\big) \,.
\end{equation} 
The time evolution induced by the Hamiltonian $\fh^{(2)}$ is given by
\begin{align}
\dot p_k &= \{\fh^{(2)}\,,\,p_k\}_{(2)} = \frac{\nu}{4} \sum_{i,j=1}^N p_i p_j p_k \,\fs_{ij}\fs_{jk}\fs_{ik}\,(\cS_{ik}-\cS_{jk}) \,, \\
\dot q_{kl} &= \{\fh^{(2)}\,,\,q_{kl}\}_{(2)} \nonumber \\ 
&= \frac{1}{4} \sum_{i,j=1}^N p_i p_j \,\fs_{ij} \Big( \fs_{jk}\cQ_{ik} - \fs_{ik}\cQ_{jk} - \fs_{jl}\cQ_{il} + \fs_{il}\cQ_{jl} + (\fs_{ik} - \fs_{jk} - \fs_{il} + \fs_{jl})\cQ_{ij} \Big) \,.
\end{align}

For $N=3$, one gets
\begin{equation}
\begin{split}
\tau_3(z) &= \sum_{i,j,k=1}^N\cL_{ij}(z)\cL_{jk}(z)\cL_{ki}(z) 
= \Lambda(z)\,\fh^{(3)} + \sfrac{3}{2} \Lambda(z)^2\,\fh^{(2)}\fh^{(1)} + \Lambda(z)^3 \big(\fh^{(1)}\big)^3 \,,
\end{split}
\end{equation}
where
\begin{equation}
\begin{split}
\fh^{(3)} &= -\sfrac{1}{8}\,\sum_{i,j,k=1}^N p_ip_jp_k\,\fs_{ij}\fs_{jk}\fs_{ki} \big( \fs_{ij}\cQ_{ij} + \fs_{jk}\cQ_{jk} + \fs_{ki}\cQ_{ki} \big) \,.
\end{split}
\end{equation}
The time evolution corresponding to the Hamiltonian $\fh^{(3)}$ reads
\begin{align}
\dot p_l =\ & \{\fh^{(3)}\,,\,p_l\}_{(2)} \nonumber \\
=\ &\frac{\nu}{32}\, \sum_{i,j,k=1}^N p_i p_j p_k p_l \, \fs_{ij}\fs_{jk}\fs_{ki} \Big( 
4\big( \fs_{il} (\fs_{kl} - \fs_{jl}) \cQ_{il} + \fs_{jl} (\fs_{il} - \fs_{kl}) \cQ_{jl} + \fs_{kl} (\fs_{jl} - \fs_{il}) \cQ_{kl} \big) \nonumber \\
& + \fs_{ij} \fs_{kl} \cQ_{ij} (\cQ_{kl}-4) + \fs_{jk} \fs_{il} \cQ_{jk} (\cQ_{il}-4) + \fs_{ki} \fs_{jl} \cQ_{ki} (\cQ_{jl}-4)
\Big) \,,\label{eq:h3-p} \\
\dot q_{mn} =\ & \{\fh^{(3)}\,,\,q_{mn}\}_{(2)} = \frac{1}{32} \sum_{i,j,k=1}^N p_i p_j p_k \,\fs_{ij}\fs_{jk}\fs_{ki} \Big( \cS_{ij}(\fs_{im}\cQ_{im}-\fs_{jm}\cQ_{jm}-\fs_{in}\cQ_{in}+\fs_{jn}\cQ_{jn}) \nonumber \\
& + \cS_{jk}(\fs_{jm}\cQ_{jm}-\fs_{km}\cQ_{km}-\fs_{jn}\cQ_{jn}+\fs_{kn}\cQ_{kn}) \nonumber \\
& + \cS_{ki}(\fs_{km}\cQ_{km}-\fs_{im}\cQ_{im}-\fs_{kn}\cQ_{kn}+\fs_{in}\cQ_{in}) \nonumber \\
& -\big( \cS_{mi}-\cS_{ni} + \cS_{mj}-\cS_{nj} + \cS_{mk}-\cS_{nk} \big)\big( \fs_{ij}\cQ_{ij} + \fs_{jk}\cQ_{jk} + \fs_{ki}\cQ_{ki} \big) \,.
\label{eq:h3-q}
\end{align}
In order to simplify the expressions occurring in equations \eqref{eq:h3-p} and \eqref{eq:h3-q}, we used the following property of the sign function: $\fs_{ij}\fs_{jk}\fs_{ki} = -(\fs_{ij}+\fs_{jk}+\fs_{ki})$, and, for $i,j,k$ distinct, $\fs_{ij}\fs_{jk}+\fs_{jk}\fs_{ki}+\fs_{ki}\fs_{ij}=-1$.

One can go further, but in practice the expressions become rather cumbersome. We provide in appendix \ref{appC} the expansion of $\tau_m(z)$, 
$m=4,5,6,7,$ but refrain from writing the Hamiltonians and the corresponding time evolutions.

\subsection{General expression for the Hamiltonians\label{sect:transfer}} 
As already stated, computing directly $\tau_m(z)$ for general $m$ doesn’t seem easily accessible at the moment. 
However, we can exhibit some general properties and a conjecture that could lead to a full determination of them.
\begin{prop}
The transfer matrix $\tau_m(z)$ is a polynomial in $\Lambda(z)$ of degree $m$ with no constant term, and a degree 1 term of the form
\begin{equation}\label{eq:hm}
\begin{split}
\fh^{(m)}&= m\tr\Big((T+\sfrac{1}{2}\gamma\,A)\,(S+A)^{m-1}\Big).
\end{split}
\end{equation}
The transfer matrix $\tau_m(z)$ is also a homogeneous, symmetric function of $p_1,...,p_N$ of degree $m$.
\end{prop}
\proof
The $\Lambda$-dependence can be proved in the following way. 
As in \eqref{eq:LRB1}, one first remarks that the Lax matrix \eqref{eq:L-linear} can be decomposed as $\cL(z)=S+\Lambda(z)T+\xi(z)A$, 
where the symmetric matrices $T$ and $S$, and antisymmetric matrix $A$ are defined in \eqref{eq:A}--\eqref{eq:S}. Then, when computing $\tr(\cL(z)^m)$, only terms with an even number of $A$ matrices contribute to $\tau_m(z)$. It implies that the dependence in the spectral parameter $z$ is a polynomial in $\Lambda(z)$ and $\xi(z)^2$. By direct calculation one shows that 
\begin{equation}\label{eq:xi2}
\xi(z)^2=\Lambda(z)^2+\gamma\,\Lambda(z)+1,
\end{equation}
 which implies that we get a degree $m$ polynomial in $\Lambda(z)$.

The degree 0 term is obtained by setting $\Lambda(z)=0$, leading to $\tr\big((S+A)^m\big)$. When we have $q_1>q_2>...>q_N$, the matrix $\cA\equiv S+A$ is strictly  triangular, so that $\tr(\cA^m)=0$. 
More generally, we define $w$ as the permutation of $[1\,,\,N]$ such that $q_{w(1)}>q_{w(2)}>...>q_{w(N)}$.
Then, $\cA$ can be rewritten as $\cA=\sum_{i,j=1}^N \cA_{ij}E_{ij} = \sum_{i',j'=1}^N \cA_{i'j'}E'_{ij}$, where we have set $i'=w(i)$, $j'=w(j)$ and $E'_{ij}=E_{i'j'}$. In the $E'_{ij}$ basis, $\cA$ is strictly triangular, so that $\tr(\cA^m)=\sum_{i=1}^N(\cA^m)_{ii} = \sum_{i'=1}^N(\cA^m)_{i'i'}=0$.

To get the degree 1 term, we differentiate $\tau_m(z)$ w.r.t. $\Lambda(z)$ and then set $\Lambda(z)=0$:  
\begin{equation}\label{eq:hm2}
\begin{split}
\fh^{(m)}&=\frac{d\tau_m(z)}{d\Lambda(z)}\,\Big|_{\Lambda=0} = \sum_{k=0}^{m-1} \tr\Big(\cL(z)^k\big(T+
\frac{d\xi(z)}{d\Lambda(z)}A\big)\cL(z)^{m-1-k}\Big)\Big|_{\Lambda=0}\\
&=\sum_{k=0}^{m-1} \tr\Big((S+A)^k\,(T+\sfrac{1}{2}\gamma\,A)\,(S+A)^{m-1-k}\Big)\\
&= m\tr\Big((T+\sfrac{1}{2}\gamma\,A)\,(S+A)^{m-1}\Big).
\end{split}
\end{equation}
To get the second line in \eqref{eq:hm2}, we have inferred from \eqref{eq:xi2} the properties $\xi(z)|_{\Lambda(z)=0}=1$ and $\frac{d\xi(z)}{d\Lambda(z)}|_{\Lambda(z)=0}=\sfrac{1}{2}\gamma$.
\qed

From the  expressions of $\tau_m(z)$, $m=2,3,...,7$, we are led to postulate the following conjecture:
\begin{conj}\label{conj:taum}
The transfer matrices $\tau_m(z)$ can be expanded as 
\begin{equation}
\tau_m(z) = \sum_{j=1}^m\Lambda(z)^j\sum_{\atopn{0< k_j\leq k_{j-1}\leq...\leq k_1}{k_1+k_2+...+k_j=m}}
a_j(k_1,k_2,..,k_j)\,\fh^{(k_1)}\fh^{(k_2)}\cdots\fh^{(k_j)}\,,
\end{equation}
with $a_1(m)=1$.
 We conjecture the following values for the coefficients
 \begin{equation}
 a_m(1,1,...,1)=1\,,\qquad a_{j+1}(m-j,\underbrace{1,...,1}_j)=\frac{m}{m-j}\,,\ 0\leq j<m-1\,,
 \end{equation}
  the other coefficients $a_j(k_1,k_2,..,k_j)$ remain to be determined.
 \end{conj}

The time evolutions can then be deduced from the relations
\begin{eqnarray}
&&\big\{\tau_m(z_1)\,,\,\cL_2(z_2)\big\}_{(2)} = m\Big[ \tr_1\Big(\big(r_{12}(z_1,z_2)+r_{12}(z_1^{\sigma},z_2)^{t_1}\big)\cL_1(z_1)^m\Big)\,,\,\cL_2(z_2)\Big]\,,\qquad
\label{eq:taumL}\\
&&\big\{\fh^{(m)}\,,\,\cL(z)\big\}_{(2)} = m \frac{\nu}4 \Big[ H^{(m)}-\tr_1(P_{12}H_1^{(m)})\,,\, \cL(z)\Big]\,,
\label{eq:hmL}
\\
&&H^{(m)} = \sum_{j+k=m} (S+A)^j\,(T+\frac{\gamma}2A)\,(S+A)^k +(S-A)^j\,(T-\frac{\gamma}2A)\,(S-A)^k\,,\quad
\end{eqnarray}
Relation \eqref{eq:taumL} is a direct consequence of the expression of $\tau_m(z)$ and the quadratic Poisson structure \eqref{eq:rll-quad}. Relation \eqref{eq:hmL} is deduced from \eqref{eq:taumL} taking the derivative w.r.t. $\Lambda(z)$, and noticing that $\Lambda(z)=0$ corresponds to $z=\pm1$. We remind that $P$ is the signed permutation, see \eqref{eq:Psigned}.

\subsection{Bi-Hamiltonian structure}
Since the $r$-matrix $\fr_{12}(z_1,z_2)$ in the definition of $\{.,.\}_{(1)}$, see \eqref{eq:PBspec}, is not antisymmetric, one cannot use the results of \cite{LiParm} to prove the bi-Hamiltonian structure of the model. However, the Poisson brackets $\{.,.\}_{(1)}$ and $\{.,.\}_{(2)}$ are compatible in the sense that the bilinear mapping $\{.,.\}_{\eta}=\{.,.\}_{(1)}+\eta\,\{.,.\}_{(2)}$ also defines a Poisson structure for all values of $\eta$. To prove it, one has to check directly the Poisson structure of $\{.,.\}_{\eta}$. It  is clear that $\{.,.\}_{\eta}$ is anti-symmetric. One has then to check the Jacobi identity.
It amounts to show that
\begin{equation}\label{eq:Jaco-y}
\begin{split}
&\big\{\{A,B\}_{(2)},C\big\}_{(1)}+ \big\{\{B,C\}_{(2)},A\big\}_{(1)}+ \big\{\{C,A\}_{(2)},B\big\}_{(1)}\\
&+ \big\{\{A,B\}_{(1)},C\big\}_{(2)}+ \big\{\{B,C\}_{(1)},A\big\}_{(2)}+ \big\{\{C,A\}_{(1)},B\big\}_{(2)}=0
\end{split}
\end{equation}
for arbitrary functions $A,B,C$ on the phase space. It is sufficient to check that the relation \eqref{eq:Jaco-y} is obeyed for the variables $p_k$ and $q_{i,i+1}$. We checked that for $A,B,C\in \{p_1,...,p_N,q_{12},q_{23},...,q_{N-1,N}\}$, relation \eqref{eq:Jaco-y} is indeed obeyed.

\paragraph{Lenard duality relations.}
From the explicit expressions of the Hamiltonians $\fh^{(m)}$ given above, one can check that the bi-Hamiltonian structure obeys the Lenard duality relations \cite{K-Schwarzb-Magri}
\begin{equation}
\begin{split}
& \big\{\fh^{(2)}\,,\,p_k\big\}_{(1)}=2\,\big\{\fh^{(1)}\,,\,p_k\big\}_{(2)}\,,\qquad  \big\{\fh^{(2)}\,,\,q_{ij}\big\}_{(1)}=2\,\big\{\fh^{(1)}\,,\,q_{ij}\big\}_{(2)}\,, \\
& \big\{\fh^{(3)}\,,\,p_k\big\}_{(1)}=\frac32\,\big\{\fh^{(2)}\,,\,p_k\big\}_{(2)}\,,\qquad  \big\{\fh^{(3)}\,,\,q_{ij}\big\}_{(1)}=\frac32\,\big\{\fh^{(2)}\,,\,q_{ij}\big\}_{(2)}\,, \\ 
& \big\{\fh^{(4)}\,,\,p_k\big\}_{(1)}=\frac43\,\big\{\fh^{(3)}\,,\,p_k\big\}_{(2)}\,,\qquad  \big\{\fh^{(4)}\,,\,q_{ij}\big\}_{(1)}=\frac43\,\big\{\fh^{(3)}\,,\,q_{ij}\big\}_{(2)}\,.
\end{split}
\end{equation}
These relations lead us to conjecture the Lenard duality form
\begin{equation}
\begin{split}
&\big\{\frac{\fh^{(m+1)}}{m+1}\,,\,p_k\big\}_{(1)}=\big\{\frac{\fh^{(m)}}{m}\,,\,p_k\big\}_{(2)}\,,\quad  
\big\{\frac{\fh^{(m+1)}}{m+1}\,,\,q_{ij}\big\}_{(1)}=\,\big\{\frac{\fh^{(m)}}{m}\,,\,q_{ij}\big\}_{(2)}\,.
\end{split}
\end{equation}

\subsection{Cyclic variable\label{sect:cycl}} 
As we already mentioned (see remark \ref{rmk:cyclic}), we know that there should be a cyclic variable that Poisson-commutes with all variables defining the phase space. 
From the conjecture \ref{conj:taum} one can argue that the cyclic variable mentioned in the remark \ref{rmk:cyclic} must be a polynomial in $\fh^{(1)}, \fh^{(2)}, ...,\fh^{(N)}$. 
Indeed, since $\cL(z)$ is a $N\times N$ matrix whose entries $\cL_{ij}(z)$ depend solely on the variables $q_{ij}$, $p_i$ and $p_j$  (and not on the other $q_{kl}$ and $p_k$ variables), the matrix $\cL(z)$ can be viewed as completely generic. Then, the traces of its powers $\tau_m(z)=\tr\big(\cL(z)^m\big)$, $m\leq N$ are all independent. This in turn implies that the Hamiltonians $\fh^{(1)}, \fh^{(2)}, ...,\fh^{(N)}$ are all independent. However, this generates a $N$-dimensional Poisson-commutative algebra for a phase space of dimension $2N-1$. This is possible only if  the cyclic variable belongs to this abelian algebra. Indeed we have
\begin{prop}\label{prop:cycl}
Let $w$ be the permutation such that $q_{w(1)}>q_{w(2)}>...>q_{w(N)}$. 
\begin{enumerate}
\item The Hamiltonian $\fh^{(N)}$ takes the form
\begin{equation}
\frac{\fh^{(N)}}{N2^{N-1}}= \Big(\prod_{i=1}^N p_i\Big)\Big(\cosh(\frac{\nu}2q_{w(N),w(1)})+\frac{\gamma}2\,\sinh(\frac{\nu}2q_{w(N),w(1)})\Big)\prod_{i=1}^{N-1} \sinh(\frac{\nu}2q_{w(i),w(i+1)})\,.
\label{eq:hN}
\end{equation}
\item $\fh^{(N)}$ is cyclic: $\{p_i\,,\,\fh^{(N)}\}=0$ and $\{q_{ij}\,,\,\fh^{(N)}\}=0$, $\forall i,j=1,...,N$.
\end{enumerate}
\end{prop}
\proof
We start by proving the property in the case $w=id$, that is to say $q_1>q_2>...>q_N$. In this case the matrix $(S+A)$ is strictly triangular, and $(S+A)^{N-1}$ has only one non-zero entry, in position $(1,N)$.  This  entry reads
\begin{equation}
2^{N-1}\Big(\prod_{i=2}^{N-1}p_i\Big)\sqrt{p_1p_N}\prod_{i=1}^{N-1} \sinh(\frac{\nu}2q_{i,i+1})\,.
\end{equation}
Then, from expression \eqref{eq:hm2}, we get the statement \eqref{eq:hN} for $w=id$.

Next, one has to prove the cyclicity of this expression. 
Before computing the Poisson brackets $\{p_i\,,\,\fh^{(N)}\}$ and $\{q_{i,i+1}\,,\,\fh^{(N)}\}$, we prepare some notations. Let us set $\gamma=2\coth\theta$, so that the Hamiltonian $\fh^{(N)}$ writes, for $w=id$, 
\begin{equation}
\fh^{(N)}= N2^{N-1}\Big(\prod_{i=1}^N p_i\Big)\Big(\prod_{i=1}^{N-1} \sinh(\frac{\nu}2q_{i,i+1})\Big)\;\frac{\sinh(\frac{\nu}2q_{N1}+\theta)}{\sinh\theta}\,.
\label{eq:hNbis}
\end{equation}
Moreover the quantities $\cQ_{ik}$ and $\cS_{ik}$ take the form
\begin{equation}\label{eq:SQbis}
\frac14\,\cQ_{ik} = \frac{\sinh({\nu}|q_{ik}|-\theta)}{\sinh\theta} + 1
\quad \text{and} \quad
\frac14\,\cS_{ik} = \frac{\cosh({\nu}|q_{ik}|-\theta)}{\sinh\theta} \,.
\end{equation}
We first consider the Poisson bracket $\{p_i\,,\,\fh^{(N)}\}$. A direct calculation leads to
\begin{equation}\label{eq:pihN}
\frac{\{p_i\,,\,\fh^{(N)}\}}{\fh^{(N)}} = 
\underbrace{\sum_{k=1}^N \frac1{p_k}\,\{p_i,p_k\}}_{\cI_1}  
+ \underbrace{\frac{\nu}2\sum_{k=1}^{N-1}\coth(\frac{\nu}2q_{k,k+1})\{p_i,q_{k,k+1}\}}_{\cI_2} 
+ \underbrace{\coth(\frac{\nu}2q_{N1}+\theta)\{p_i,q_{N1}\}}_{\cI_3}\,.
\end{equation}
Given the Poisson brackets \eqref{eq:qijpk} and the expressions \eqref{eq:SQbis}, one gets, up to a global factor $\frac12\,\nu p_i$,
\begin{align}
\cI_1 &
= 2\sum_{k=1}^N \fs_{ik} \frac{\sinh({\nu}|q_{ik}|-\theta)}{\sinh\theta} \,, \\
\cI_2 
&= \sum_{k=1}^{N-1} \coth(\frac{\nu}2q_{k,k+1}) \Big(\frac{\cosh({\nu}|q_{ik}|-\theta)}{\sinh\theta} - \frac{\cosh({\nu}|q_{i,k+1}|-\theta)}{\sinh\theta}\Big) \,,\\
\cI_3 &
= \coth(\frac{\nu}2q_{N1+\theta}) \Big(\frac{\cosh({\nu}q_{iN}-\theta)}{\sinh\theta} - \frac{\cosh({\nu}q_{i1}+\theta)}{\sinh\theta}\Big) \,.
\end{align}
Using the following formula for any indices $a,b,c$,
\begin{equation}
\coth(\frac{\nu}2q_{ab}) \big(\cosh({\nu}q_{ca}\pm\theta) - \cosh({\nu}q_{cb}\pm\theta)\big)
= -\big( \sinh({\nu}q_{ca}\pm\theta) + \sinh({\nu}q_{cb}\pm\theta) \big)\,,
\end{equation}
the term $\cI_2$ is transformed as (taking care of the absolute value $|q_{ik}|$)
\begin{equation}
\begin{split}
\cI_2 &= \frac{-1}{\sinh\theta}\Big(
\sum_{k=1}^{i-1} \sinh({\nu}q_{ik}+\theta) + \sinh({\nu}q_{i,k+1}+\theta)
+ \sum_{k=i}^{N-1} \sinh({\nu}q_{ik}-\theta) + \sinh({\nu}q_{i,k+1}-\theta)
\Big)  \\
&= \frac{-2}{\sinh\theta}\Big(
\sum_{k=1}^{i-1} \sinh({\nu}q_{ik}+\theta) + \sum_{k=i+1}^{N} \sinh({\nu}q_{ik}-\theta)
- \frac{1}{2}\,\sinh({\nu}q_{i1}+\theta) - \frac{1}{2}\,\sinh({\nu}q_{iN}-\theta) \Big) 
\,.
\end{split}
\end{equation}
In the same way, the term $\cI_3$ is transformed using the formula 
\begin{equation}\label{eq:formulN1}
\coth(\frac{\nu}2q_{N1}+\theta) \big(\cosh({\nu}q_{iN}-\theta) - \cosh({\nu}q_{i1}+\theta)\big)
= -\big( \sinh({\nu}q_{iN}-\theta) + \sinh({\nu}q_{i1}+\theta) \big)\,.
\end{equation}
Then, adding the three terms altogether one gets $\cI_1+\cI_2+\cI_3=0$, which proves that $p_i$ Poisson commutes with the Hamiltonian $\fh^{(N)}$ for all $i$.

Similarly, we have 
\begin{equation}\label{eq:qii1hN}
\begin{split}
&\dfrac{\{q_{i,i+1}\,,\,\fh^{(N)}\}}{\fh^{(N)}} =\\
&=\underbrace{\sum_{k=1}^N \frac1{p_k}\,\{q_{i,i+1},p_k\}}_{\cI'_1}  
+ \underbrace{\frac{\nu}2\sum_{k=1}^{N-1}\coth(\frac{\nu}2q_{k,k+1})\{q_{i,i+1},q_{k,k+1}\}}_{\cI'_2} 
+ \underbrace{\coth(\frac{\nu}2q_{N1}+\theta)\{q_{i,i+1},q_{N1}\}}_{\cI'_3}\,.
\end{split}
\end{equation}
Again, using the expressions \eqref{eq:SQbis}, the following formula for any indices $a,b,c$,
\begin{equation}
\coth(\frac{\nu}2q_{ab}) \big(\sinh({\nu}q_{ca}\pm\theta) - \sinh({\nu}q_{cb}\pm\theta)\big)
= -\big( \cosh({\nu}q_{ca}\pm\theta) + \cosh({\nu}q_{cb}\pm\theta) \big)\,,
\end{equation}
and a formula similar to \eqref{eq:formulN1} for the term $\cI'_3$, one is led to $\cI'_1+\cI'_2+\cI'_3=0$ for generic $1<i<N-1$. Note that the term $\cI'_2$ has to be computed with care when dealing with the particular values $k=i\pm1$ in the sum ($k=i$ gives obviously zero).
The cases $i=1$ and $i=N-1$ are treated analogously, taking into account that some terms are then vanishing. Hence the variables $q_{i,i+1}$ Poisson commute with the Hamiltonian $\fh^{(N)}$ for all $1  \leq i<N$.

Thus, we have shown that $\fh^{(N)}$ is cyclic when $w=id$. It remains to consider the case of a generic $w$.

We first consider the matrix 
\begin{equation}\label{eq:S+A}
S+A\equiv \sum_{i,j=1}^N u(p_i,p_j,q_{ij})\, E_{ij}\,,\quad\mbox{where}\quad u(p,\bar p,q)=\sqrt{p\bar p} \,(1+\sgn(q))\,\sinh \frac{\nu}{2}q.
\end{equation}
As already stated, when $w=id$ (i.e. $q_{ij}>0\ \Leftrightarrow\ i<j$), the matrix $S+A$ is strictly triangular and $(S+A)^{N-1}$ has only one non-zero entry at position $(1,N)$, of the form 
$\prod_{i=1}^{N-1} u(p_i,p_{i+1},q_{i,i+1})$. Now, for generic $w$, we set $i'=w(i)$, $j'=w(j)$, $p'_i=p_{w(i)}$, $q'_{ij}=q_{w(i)w(j)}$ and $E'_{ij}=E_{i'j'}$  in \eqref{eq:S+A}. We have $\sgn(q'_{ij})>0$ iff $i'<j'$, so that $S+A$ is strictly triangular in the $E'_{ij}$ basis. Hence, the only non-zero entry of 
$(S+A)^{N-1}$ is at position $(1',N')$ and takes the value $\prod_{i=1}^{N-1} u(p'_i,p'_{i+1},q'_{i,i+1})$. This leads to the final expression \eqref{eq:hN}.

Finally, the expression \eqref{eq:hN} for generic $w$ can be brought to the same form as $w=id$ using the variables 
$q'_{ij}$ and $p'_i$. Since the Poisson brackets \eqref{eq:qijpk} keep the same form in 
$q'_{ij}$ and $p'_i$ variables, this shows that the proof of cyclicity for $w=id$ implies the proof of cyclicity for all $w$.
This ends the proof of property \ref{prop:cycl}.
\qed

\section{Conclusion\label{sec:conclu}}
We have now established a very rich picture of integrability structures from the original Camassa Holm peakon dynamics. We had established in \cite{AFR23} the existence of a two-parameter extension 
(with a deformation $\gamma$ and a spectral parameter $z$) of the original  Lax matrix and $r$-matrix structure. We have constructed here a consistent bi-Hamiltonian structure for this extension. Different limiting procedures lead back to the original peakon model or to its Ragnisco--Bruschi extension. 
Concerning the bi-Hamiltonian structure of the model, it would be desirable to prove the conjecture we have postulated on the expansion of the transfer matrices. An explicit expression for the Hamiltonians $\fh^{(m)}$, going beyond the expression \eqref{eq:hm} would be also interesting. It may proceed along lines similar to those used in the computation of the Hamiltonian $\fh^{(N)}$, based on the triangular structure of the powers $(S+A)^m$.

It is natural to study the existence of such extensions for other peakon dynamics and a relevant candidate is the peakon dynamics associated to Degasperis--Procesi equation \cite{dGHH,DGP}. The situation at this time is very different: expressed in our language one only knows a single non-parametric Lax matrix (with no deformation or spectral parameter) for a single quadratic Poisson structure \cite{AFR22}. Looking for similar extensions in this case may also help to understand the difficulty in building a linear Poisson structure for Degasperis--Procesi peakons. Indeed, it is impossible to get a linear Poisson structure directly from the Poisson structures of the original fluid equation. This suggests the occurence of singular behaviors in what would correspond to the limits described in Section \ref{sec:mat-r}.

It seems also interesting to look more closely at the $r$-matrix \eqref{eq:Rsplity} found in the halved algebra and find new integrable models described by this interesting mixture of Toda (the $\Gamma$ term) and maximal-coadjoint orbits (the $\Pi$ term)  $r$-matrices. 

\subsubsection*{Acknowledgments}
We acknowledge financial support from the USMB AAP grant \textsl{POSITIPh}.
J.A. wishes to thank the LAPTh for financial support and warm hospitality during the course of this investigation.

\appendix

\section{Hamiltonian dynamics for the halved algebra\label{app:hamil}}
\subsection{First Hamiltonian structure}
We  look at the first Hamiltonian structure associated to the Lax matrix $\ell(z)$.
Indeed, from the relation \eqref{eq:PB0ell}, we know that 
\begin{equation}
\begin{split}
&\{\ft^{(m_1)}(z_1)\,,\,\ft^{(m_2)}(z_2)\}_{(1)} =0\,,\quad \forall\,z_1,z_2,m_1,m_2\,,\\
&\ft^{(m)}(z)=\text{tr}\Big(\ell(z)^m\Big)\,.
\end{split}
\end{equation}
Upon expansion in $z$, $\ft^{(m)}(z)$ generate an infinite number of conserved quantities, indexed by $m$. Introducing 
\begin{equation}
\fp=(\sqrt{p_1},...,\sqrt{p_n})\,,\quad \pp=p_1+...+p_n\,,\quad \Sigma=\sum_{i,j=1}^N\fs_{ij}\,E_{ij}
\end{equation}
and the expression \eqref{eq:Lsplity2} for $\ell(z)$, one gets
\begin{equation}\label{eq:t(z)split}
\begin{split}
\ft^{(m)}(z)& =\frac1{2^m}\,\sum_{k_1,...,k_m=1}^N p_{k_1}\cdots p_{k_m}\,(z+\fs_{k_1k_2})(z+\fs_{k_2k_3})\cdots(z+\fs_{k_mk_1}) \\
&=\pp^{m-1}\, \sum_{l=0}^{[m/2]} \begin{pmatrix}m\\ 2l \end{pmatrix}z^{m-2l}\,(\fp \Sigma^{2l}\fp^t )
\end{split}
\end{equation}
which leads to the commuting Hamiltonians
\begin{equation}\label{eq:Hamsplit}
\begin{split}
&\fh^{(m)}_{l}= \pp^{m-1}\, (\fp \Sigma^{2l}\fp^t )\,,\quad0\leq l\leq \left[\frac{m}2\right]\,,\quad1\leq m\leq N\,,
\\
&\big\{\fh^{(m_1)}_{l_1}\,,\,\fh^{(m_2)}_{l_2}\big\}_{(1)}=0\,,\quad \forall\, l_1,l_2,m_1,m_2\,.
\end{split}
\end{equation}
In particular,
\begin{equation}
\begin{split}
\fh^{(m)}_{0} &= \pp^{m}\, ,\\
\fh^{(m)}_{1} &= (1-n)\pp^{m}  +2\pp^{m-1}\,\sum_{1\leq i< j\leq N} (2j-2i-N)\,\sqrt{p_ip_{j}}\\
&= \pp^{m}  +\pp^{m-1}\,\sum_{i=1}^N\sum_{j=1}^{N} (2|j-i|-N)\,\sqrt{p_ip_{j}}\,.
\end{split}
\end{equation}
Remark that we have
\begin{equation}\label{eq:hmhl}
\fh^{(m)}_{l}= \fh^{(m-1)}_0\,\wt\fh_l+\fh^{(m)}_0\,,\quad \wt\fh_l=\fp\, (\Sigma^{2l}-\II)\,\fp^t\,,\quad0\leq l\leq \left[\frac{m}2\right]\,,\quad1\leq m\leq N\,.
\end{equation}
Hence, the study of the Hamiltonians $\fh^{(m)}_{l}$ can be reduced to the study of the Hamiltonians $\fh^{(m)}_{0}$ and $\wt\fh_{l}$.

\subsection{Time evolutions}
The time evolution induced by any of  the Hamiltonians $\fh^{(m)}_{0}$ is rather trivial:
\begin{equation}
\begin{split}
\frac{d p_k }{dt_m} &= \{\fh^{(m)}_{0}\,,\,p_k\}_{(1)}=0\,,\qquad
\frac{d q_k }{dt_m} = \{\fh^{(m)}_{0}\,,\,q_k\}_{(1)} =m\pp^{m-1}\,,
\end{split}
\end{equation}
where the Poisson brackets have been computed using \eqref{eq:PBcano}. They all lead to a uniform displacement, identical for all particles
\begin{equation}
\begin{split}
p_k (t_m) &= p_k(0)\,,\qquad
q_k(t_m) = q_k(0)+v_m t_m\,,\quad v_m=m\pp^{m-1}\,.
\end{split}
\end{equation}

The time evolution corresponding to the Hamiltonian 
\begin{equation}
\wt \fh_{1}=\frac{\fh^{(m)}_{1}-\fh^{(m)}_{0}}{\fh^{(m-1)}_{0}}=\sum_{1\leq i,j\leq N} (2|j-i|-N)\,\sqrt{p_ip_{j}}
\end{equation}
is more involved
\begin{equation}
\begin{split}
\dot{p_k } &= \{\wt\fh_{1}\,,\,p_k\}_{(1)}=0\,,\qquad
  \dot{q_k } 
 =\frac12\,\sum_{j=1}^N (2|j-k|-N)\,\sqrt{\frac{p_{j}}{p_k}}\,.
\end{split}
\end{equation}
The displacement is still uniform, but each particle has its own speed
\begin{equation}
\begin{split}
p_k (t) &= p_k(0)\,,\qquad
q_k(t) = q_k(0)+v_kt\,,\quad v_k=\frac12\,\sum_{j=1}^N (2|k-j|-n)\,\sqrt{\frac{p_{j}}{p_k}}\,.
\end{split}
\end{equation}
More generally, the time evolutions associated to the Hamiltonians $\wt\fh_{l}$ take the form
\begin{equation}
\frac{d p_k }{dt_{l}} = \{\wt\fh_{l}\,,\,p_k\}_{(1)}=0\,,\qquad
\frac{d q_k }{dt_{l}} = \{\wt\fh_{l}\,,\,q_k\}_{(1)} =\frac{d}{dp_k}\wt\fh_{l}\,.
\end{equation}

However, since the Hamiltonians are always polynomials in $p_k$, $k=1,2,...,N$ only, each momentum $p_j$ is  conserved. 
It is only through the twisting procedure that one is led to non conserved momenta
obeying  a peakon dynamics.

\subsection{Second Hamiltonian structure\label{app:Ham2-halved}}
Now, we start from the second Poisson structure of the halved algebra \eqref{eq:PB2ell}. 
We can still use the transfer matrices \eqref{eq:t(z)split} of the first Poisson structure,
since they obey
\begin{equation}
\begin{split}
\big\{\ft^{(m_1)}(z_1)\,,\, \ft^{(m_2)}(z_2)\big\}_{(2)}  =\ & \text{tr}_1\text{tr}_2\big\{\ell_1(z_1)^{m_1}\,,\, \ell_2(z_2)^{m_2}\big\}_{(2)} \\
=\ & m_1m_2\,\text{tr}_{12}\, \ell_1(z_1)^{m_1-1}\, \ell_2(z_2)^{m_2-1}\,\big\{\ell_1(z_1)\,,\, \ell_2(z_2)\big\}_{(2)} \\
=\ & m_1m_2\,\text{tr}_{12}\,\ell_1(z_1)^{m_1-1}\, \ell_2(z_2)^{m_2-1}\,\big[r_{12}(z_1,z_2)\,,\,\ell_1(z_1)\,\ell_2(z_2)\big] \\
=\ & \text{tr}_{12}\,\big[r_{12}(z_1,z_2)\,,\,\ell_1(z_1)^{m_1}\,\ell_2(z_2)^{m_2}\big]=0\,.
\end{split}
\end{equation}
Now, we choose for the Lax matrix the expression \eqref{eq:Lsplity}-\eqref{eq:Lsplity2}.
The second Poisson structure  provides
\begin{equation}\label{eq:PBell}
\big\{p_i\,,\,p_j\big\}_{(2)} =0\,,\qquad \big\{q_i\,,\,p_j\big\}_{(2)} =\frac12p_i\,\delta_{i,j}\,,\qquad \big\{q_i\,,\,q_j\big\}_{(2)} =-\frac1{\nu}\,\fs_{ij}\,. 
\end{equation}
Then, the Hamiltonians \eqref{eq:hmhl} lead to the  time evolution 
\begin{equation}
\frac{dp_i}{dt_{m,l}}=0\quad\text{and}\quad \frac{dq_i}{dt_{m,l}}=\big\{\fh^{(m)}_{l}\,,\,q_i\big\}_{(2)}=-\frac12p_i\,\frac{d}{dp_i}\fh^{(m)}_{l}\,.
\end{equation}

\paragraph{Bi-Hamiltonian structure.} Since the Poisson structures $\{.\,,\,.\}_{(1)}$ and $\{.\,,\,.\}_{(2)}$, see \eqref{eq:PB0ell} and \eqref{eq:PB2ell}, are induced from the antisymmetric $r$-matrix \eqref{eq:rmat-split}, one deduces \cite{LiParm} that 
 the two Poisson structures are compatible in the sense that 
$a\,\{.\,,\,.\}_{(1)}+b\,\{.\,,\,.\}_{(2)}$ is also a Poisson bracket for any complex numbers $a$ and $b$. 

However,  the time evolutions associated to the different Hamiltonians and Poisson structures do not obey a Lenard duality relation \cite{K-Schwarzb-Magri}, but the relation 
\begin{equation}
\big\{\fh^{(m)}_{l}\,,\,q_i\big\}_{(2)}=-\frac12\,\big\{\fh^{(m)}_{l}\,,\,p_iq_i\big\}_{(1)}\,.
\end{equation}

\section{Peakon Poisson brackets inherited from the Lax matrix\label{appB}}
We present here the general strategy to compute the Poisson brackets of $q_i$ and $p_k$ variables from the Poisson brackets \eqref{eq:PB-quad} of the Lax matrix $L(z)$.

We consider a form
\begin{equation}\label{solu-L}
\begin{split}
L_{ij}(z)=&\sqrt{p_ip_j}\,\Big(h_{ij}^+(z)e^{\frac{\nu}2|q_i-q_j|} +h_{ij}^0(z)+h_{ij}^-(z)e^{-\frac{\nu}2|q_i-q_j|}\Big)\,,
\end{split}
\end{equation}
with arbitrary functions $h_{ij}^{\eps}(z)$, $\eps=0,\pm$.
The Lax presentation for the quadratic structure of Camassa--Holm peakons is consistently presented in this form, with  $h_{ij}^+(z)=h_{ij}^0(z)=0$ and $h_{ij}^-(z)=1$.

From the form \eqref{solu-L}, we obtain
\begin{equation}\label{eq:LHS2}
\begin{split}
&\frac{\text{l.h.s.}\eqref{eq:PB-quad}}{\sqrt{p_ip_jp_kp_l}}
\ =\  \fn'_{ij}(z_1)\fn'_{kl}(z_2)\,\{q_{ij}\,,\,q_{kl}\}_{(2)} 
\\
&
+\fn'_{ij}(z_1)\fn_{kl}(z_2)\,\Big(\frac{\{q_{ij}\,,\,p_{k}\}_{(2)}}{2p_k}+\frac{\{q_{ij}\,,\,p_{l}\}_{(2)}}{2p_l}\Big)
+\fn_{ij}(z_1)\fn'_{kl}(z_2)\,\Big(\frac{\{p_{i}\,,\,q_{kl}\}_{(2)}}{2p_i}+\frac{\{p_{j}\,,\,q_{kl}\}_{(2)}}{2p_j} \Big)\\
&+\fn_{ij}(z_1)\fn_{kl}(z_2)\,\Big(\frac{\{p_{i}\,,\,p_{k}\}_{(2)}}{4p_ip_k} +\frac{\{p_{j}\,,\,p_{k}\}_{(2)}}{4p_jp_k} +
\frac{\{p_{i}\,,\,p_{l}\}_{(2)}}{4p_ip_l} +\frac{\{p_{j}\,,\,p_{l}\}_{(2)}}{4p_jp_l} \Big)
\end{split}
\end{equation}
and
\begin{equation}\label{eq:RHS2}
\begin{split}
\frac{\text{r.h.s.}\eqref{eq:PB-quad}}{\sqrt{p_ip_jp_kp_l}}
= &\ \frac{\nu}4\,\Big( \frac{z_1z_2-1}{z_1-z_2} - \fs_{ik}\Big) \fn_{kj}(z_1)\,\fn_{il}(z_2)
-\frac{\nu}4\,\Big( \frac{z_1z_2-1}{z_1-z_2} - \fs_{lj}\Big)\fn_{kj}(z_2)\,\fn_{il}(z_1)
\\
&+\frac{\nu}4\,\Big( \frac{z_1^\sigma z_2-1}{z_1^\sigma-z_2} - \fs_{jk}\Big)
\fn_{ik}(z_1)\,\fn_{jl}(z_2)
-\frac{\nu}4\,\Big( \frac{z_1^\sigma z_2-1}{z_1^\sigma-z_2} - \fs_{li}\Big)
\fn_{ki}(z_2)\,\fn_{lj}(z_1)\,,
\end{split}
\end{equation}
where 
\begin{equation}
\begin{split}
\fn_{ij}(z)=&\sum_{\eps=0,\pm1}h_{ij}^{\eps}(z)\,e^{\frac{\nu}2\eps |q_{i}-q_{j}|}\,,\\
\fn'_{ij}(z)=&\ \fs_{ij}\frac{\nu}2\Big(h_{ij}^{+}(z)\,e^{\frac{\nu}2 |q_{i}-q_{j}|} -h_{ij}^{-}(z)\,e^{-\frac{\nu}2 |q_{i}-q_{j}|}\Big)
=\frac{d}{dq_{ij}} \fn_{ij}(z)\,.
\end{split}
\end{equation}
The general strategy to find the Lax matrix and the Poisson brackets, is to first consider $\eqref{eq:LHS2} = \eqref{eq:RHS2}$ for  the case $i=j\neq k=l$, which specializes to the Poisson bracket 
$\{p_i,p_k\}$: 
\begin{equation}\label{eq:pp}
\begin{split}
\frac{\{p_{i}\,,\,p_{k}\}_{(2)}}{p_ip_k}\fn_{ii}(z_1)\,\fn_{kk}(z_2)
 =\ &
\frac{\nu}4\,\fs_{ki}\Big(\fn_{ki}(z_1)+\fn_{ik}(z_1)\Big)\Big(\fn_{ki}(z_2)+\fn_{ik}(z_2)\Big)
\\
&+\frac{\nu}4\,\frac{z_1z_2-1}{z_1-z_2}\Big(\fn_{ki}(z_1)\,\fn_{ik}(z_2)- \fn_{ki}(z_2)\,\fn_{ik}(z_1)\Big)\\
&+\frac{\nu}4\,\frac{z_1^{\sigma} z_2-1}{z_1^{\sigma}-z_2}\Big(\fn_{ik}(z_1)\,\fn_{ik}(z_2)- \fn_{ki}(z_2)\,\fn_{ki}(z_1)\Big)\,.
\end{split}
\end{equation}
Then, once knowing $\{p_i,p_k\}$, the case $i\neq j$ and $k=l$ allows to specialize to $\{q_{ij},p_k\}$, where $q_{ij}$ stand for $q_{i}-q_{j}$: 
\begin{equation}
\begin{split}\label{eq:qp}
& \fn'_{ij}(z_1)\,\fn_{kk}(z_2)
 \frac{\{q_{ij}\,,\,p_{k}\}_{(2)}}{p_k}+
\fn_{ij}(z_1)\,\fn_{kk}(z_2)\frac1{2}\Big(\frac{\{p_{i}\,,\,p_{k}\}_{(2)}}{p_ip_k} +\frac{\{p_{j}\,,\,p_{k}\}_{(2)}}{p_jp_k} \Big)
\\
&\qquad= 
\frac{\nu}4\,\Big( \frac{z_1z_2-1}{z_1-z_2} - \fs_{ik}\Big) \fn_{kj}(z_1)\,\fn_{ik}(z_2)\,
 -\frac{\nu}4\,\Big( \frac{z_1z_2-1}{z_1-z_2} 
- \fs_{kj}\Big)\fn_{kj}(z_2)\,\fn_{ik}(z_1)
\\
&\qquad\quad+\frac{\nu}4\,\Big( \frac{z_1^\sigma z_2-1}{z_1^\sigma-z_2} - \fs_{jk}\Big)
\fn_{ik}(z_1)\,\fn_{jk}(z_2)\,
 -\frac{\nu}4\,\Big( \frac{z_1^\sigma z_2-1}{z_1^\sigma-z_2} - \fs_{ki}\Big)
\fn_{ki}(z_2)\,\fn_{kj}(z_1)
\,.
\end{split}
\end{equation}
Finally, $\{q_{ij},q_{kl}\}$ is obtained in the general case $\eqref{eq:LHS2} = \eqref{eq:RHS2}$, once knowing the two Poisson brackets $\{p_i,p_k\}$ and $\{q_{ij},p_k\}$.

\section{Expressions of higher transfer matrices \label{appC}}
We provide here explicit formulas for the transfer matrices $\tau_m(z)=\tr\big(\cL(z)^m\big)$ studied in section \ref{sect:transfer}, for $m=4,5,6,7$:
\begin{equation}
\begin{split}
\tau_4(z) &= \Lambda(z)\,\fh^{(4)} + \Lambda(z)^2\,\Big(\sfrac{4}{3}\,\fh^{(3)}\fh^{(1)}+\sfrac{1}{2}\,\big(\fh^{(2)}\big)^2\Big) + 2\Lambda(z)^3\,\fh^{(2)}\big(\fh^{(1)}\big)^2 + \Lambda(z)^4 \big(\fh^{(1)}\big)^4 \,,
\end{split}
\end{equation}

\begin{equation}
\begin{split}
\tau_5(z) &= \Lambda(z)\,\fh^{(5)} + \Lambda(z)^2\,\Big(\sfrac{5}{4}\,\fh^{(4)}\fh^{(1)}+\sfrac{5}{6}\,\fh^{(3)}\fh^{(2)}\Big) + \Lambda(z)^3\,\Big(\sfrac{5}{3}\,\fh^{(3)}\big(\fh^{(1)}\big)^2 + \sfrac{5}{4}\big(\fh^{(2)}\big)^2\fh^{(1)} \Big) \\
&+ \sfrac{5}{2}\,\Lambda(z)^4\,\fh^{(2)}\big(\fh^{(1)}\big)^3 + \Lambda(z)^5 \big(\fh^{(1)}\big)^5 \,,
\end{split}
\end{equation}

\begin{equation}
\begin{split}
\tau_6(z) &= \Lambda(z)\,\fh^{(6)} + \Lambda(z)^2\,\Big(\sfrac{6}{5}\,\fh^{(5)}\fh^{(1)}+\sfrac{3}{4}\,\fh^{(4)}\fh^{(2)}+\sfrac{1}{3}\big(\fh^{(3)}\big)^2\Big) \\
&+ \Lambda(z)^3\,\Big(\sfrac{3}{2}\,\fh^{(4)}\big(\fh^{(1)}\big)^2 + 2\fh^{(3)}\fh^{(2)}\fh^{(1)} + \sfrac{1}{4}\,\big(\fh^{(2)}\big)^3 \Big) \\
&+ \Lambda(z)^4\,\Big( 2\fh^{(3)}\big(\fh^{(1)}\big)^3 + \sfrac{9}{4}\big(\fh^{(2)}\big)^2\big(\fh^{(1)}\big)^2 \Big) + 3\Lambda(z)^5\,\fh^{(2)}\big(\fh^{(1)}\big)^4 + \Lambda(z)^6 \big(\fh^{(1)}\big)^6 \,,
\end{split}
\end{equation}

\begin{equation}
\begin{split}
\tau_7(z) &= \Lambda(z)\,\fh^{(7)} + \Lambda(z)^2\,\Big(\sfrac{7}{6}\,\fh^{(6)}\fh^{(1)}+\sfrac{7}{10}\fh^{(5)}\fh^{(2)}+\sfrac{7}{12}\fh^{(4)}\fh^{(3)}\Big) \\
&+ \Lambda(z)^3\,\Big( \sfrac{7}{5}\,\fh^{(5)}\big(\fh^{(1)}\big)^2 + \sfrac{7}{4}\fh^{(4)}\fh^{(2)}\fh^{(1)} + \sfrac{7}{12}\,\fh^{(3)}\big(\fh^{(2)}\big)^2 + \sfrac{7}{9}\big(\fh^{(3)}\big)^2\fh^{(1)} \Big) \\
&+ \Lambda(z)^4\,\Big( \sfrac{7}{4}\,\fh^{(4)}\big(\fh^{(1)}\big)^3 + \sfrac{7}{8}\big(\fh^{(2)}\big)^3\fh^{(1)} + \sfrac{7}{2}\,\fh^{(3)}\fh^{(2)}\big(\fh^{(1)}\big)^2 \Big) \\
&+ \Lambda(z)^5\,\Big( \sfrac{7}{2}\,\fh^{(3)}\big(\fh^{(1)}\big)^4 + \sfrac{7}{3}\big(\fh^{(2)}\big)^2\big(\fh^{(1)}\big)^3 \Big)
+ \sfrac{7}{2}\,\Lambda(z)^6\,\fh^{(2)}\big(\fh^{(1)}\big)^5 + \Lambda(z)^7 \big(\fh^{(1)}\big)^7 \,.
\end{split}
\end{equation}

\end{document}